\documentclass[apj,iop,twocolumn]{emulateapj}
\usepackage{epstopdf}
\usepackage{CJK}
\usepackage{color}
\usepackage{pshan}
\newcommand{\beq}{\begin{equation}}
\newcommand{\eeq}{\end{equation}}
\newcommand{\bea}{\begin{eqnarray}}
\newcommand{\eea}{\end{eqnarray}}
\usepackage{bm}
\begin{document}
\def\bulletskip{0.15cm}
\begin{CJK}[HL]{KS}{}
\title{The Impact of baryonic physics on the kinetic Sunyaev-Zel'dovich Effect}

\author{Hyunbae Park\altaffilmark{1}}
\author{Marcelo A. Alvarez\altaffilmark{2,3}}
\author{J. Richard Bond\altaffilmark{2}}
\affil{$^1$Korea Astronomy and Space Science Institute, 776 Daedeok-Daero Yuseong-Gu, Daejeon, 34055, Korea}
\affil{$^2$Canadian Institute for Theoretical Astrophysics, 60 St George, Toronto ON, Canada, M5S 3H8}
\affil{$^3$Berkeley Center for Cosmological Physics, Campbell Hall 341, University of California, Berkeley CA 94720}
\shortauthors{Park, Alvarez, and Bond}


\begin{abstract}
Poorly understood ``baryonic physics" impacts our ability to predict the power spectrum of the kinetic Sunyaev-Zel'dovich (kSZ) effect. We study this in one sample high resolution simulation of galaxy formation and feedback, Illustris. The high resolution of Illustris allows us to probe the kSZ power spectrum on multipoles $\ell =10^3-3\times 10^4$.  Strong AGN feedback in Illustris nearly wipes out gas fluctuations at $k\gtrsim1~h~\rm{Mpc}^{-1}$ and at late times, likely somewhat under predicting the kSZ power generated at $z\lesssim 1$. The post-reionization kSZ power spectrum for Illustris is well-fit  by $\mathcal{D}^{z<6}_{\ell} = 1.38[\ell/3000]^{0.21}~\mu K^2$  over $3000\lesssim\ell\lesssim10000$, somewhat lower than most other reported values but consistent with the analysis of Shaw et al. Our analysis of the bias of free electrons reveals subtle effects associated with the multi-phase gas physics and stellar fractions that affect even linear scales.  In particular there are fewer electrons in biased galaxies, due to gas cooling and star formation, and this leads to an electron bias less than one even at low wavenumbers. The combination of bias and electron fraction that determines the overall suppression is relatively constant, $f_e^2b^2_{e0} \sim 0.7$, but more simulations are needed to see if this is Illustris-specific.  By separating the kSZ power into different terms, we find at least $6\, (10)\%$ of the signal at $\ell=3000\, (10000)$  comes from non-Gaussian connected four-point density and velocity correlations, $\left<\delta v \delta v\right>_{c}$, even without correcting for the Illustris simulation box size. A challenge going forward will be to accurately model long-wave velocity modes simultaneously with Illustris-like high resolution to capture the complexities of galaxy formation and its correlations with large scale flows.
\end{abstract}

 
\section{Introduction}
 
The cosmic microwave background (CMB) is a powerful probe of cosmic history through secondary anisotropies generated by free electrons in the intergalactic medium, resulting in a spectral distortion and cluster-concentrated anisotropies from inverse Compton scattering off of hot gas in galaxy clusters, the thermal Sunyaev-Zel'dovich (tSZ) effect \citep[e.g.,][]{1972CoASP...4..173S}, and spectrally undistorted temperature anisotropies from Thompson scattering from moving electrons participating in the bulk flow of the medium, the kinetic Sunyeav-Zel'dovich (kSZ) \citep[e.g.,][]{1980MNRAS.190..413S} effect. 

Of these two, the tSZ effect is much easier to detect because it is determined by electron pressure, which is much higher in clusters. The distinct spectral signature allows it to be separated from other sky signals if observations are made in multiple frequency bands. Because the tSZ effect is dominated by massive clusters, however, it gives a rather biased view of the intergalactic medium, concentrated towards the rarest peaks at relatively late times, $z\la 1.5$ \citep[e.g.,][]{2002MNRAS.336.1256K}. 
The kSZ effect, on the other hand, while much more difficult to detect, is affected by all ionized gas regardless of its temperature, providing a more global and unbiased view of the universe during and after reionization. The next generation of ground based CMB surveys, such as the Simons Observatory (SO\footnote{\url{http://simonsobservatory.org}}) and CMB-S4\footnote{\url{http://cmb-s4.org}} \citep{2016arXiv161002743A} will use the kSZ effect both as a probe of the intergalactic medium \citep{2015PhRvL.115s1301H,2016PhRvD..93h2002S, 2016PhRvL.117e1301H,2016MNRAS.461.3172S,2017JCAP...03..008D, 2017arXiv170505881B}:  cosmology with large-scale velocity flows \citep{2015ApJ...808...47M,2017JCAP...01..057S,2015PhRvD..92f3501M,2016PhRvD..94d3522A} will soon become a reality.

The kSZ effect can be detected in the cosmic microwave background alone or by cross-correlating it with tracers of large-scale structure such as galaxies and clusters. Cross correlation is done by combining CMB temperature maps with line intensity maps in the EOR \citep{2004PhRvD..70f3509C, 2005MNRAS.360.1063S, 2006ApJ...647..840A}, with projected fields of large scale structure \citep[e.g.,][]{2004ApJ...606...46D,2015PhRvD..91h3533F,2016PhRvL.117e1301H}, or by stacking on groups and clusters \citep{2016PhRvD..93h2002S,2016MNRAS.461.3172S}. These techniques use the fact that the peculiar velocities responsible for the effect are correlated with large scale density fluctuations, while the electron density fluctuations required to avoid line of sight cancellation occur on small scales. This separation of linear velocity modes on large scales and density fluctuation on small scales also greatly simplifies the modeling. 

The kSZ effect is quite challenging to detect robustly using CMB temperature measurements alone. On large scales,  the power spectrum is subdominant but has the same frequency spectrum as the much stronger primary temperature fluctuations  \citep[e.g.,][]{2002ARA&A..40..171H}. On smaller scales where it does dominate over the primary, $\ell\ga 4000$, it competes with other secondary signals such as the cosmic infrared background and tSZ, made especially difficult because they will be correlated and there are not enough  frequency bands to  separate the components in spite of their rather different frequency spectra. Experiments are in development that will help to give the resolution, sensitivity and frequency coverage to make for a very promising kSZ future in observations. 
There have been detections reported in the literature, the best so far from the South Pole Telescope at a multipole moment of $\ell \sim 3000$:  $\mathcal{D}_{\ell=3000}^{\rm kSZ}=2.9 \pm 1.3~\mu K^2$ \citep{2015ApJ...799..177G}, where $\mathcal{D}_\ell \equiv \ell(\ell + 1)\rm{C}_\ell/2\pi$. 
However, combining the South Pole Telescope, Atacama Cosmology Telescope and Planck data removes the detection and instead gives an upper bound of $\mathcal{D}_{\ell=3000}^{\rm kSZ}<1.79~\mu K^2$ \citep{2016A&A...596A.108P}.
In future CMB experiments, the kSZ power spectrum should be unveiled  by using independent polarization measurements to predict the primary temperature power spectrum and subtract it, leaving the kSZ \citep{2014JCAP...08..010C}. Another approach is beyond the 2-point power spectrum, for example correlating with the square of filtered temperature fluctuations \citep{2016arXiv160701769S}. 

One of the primary reasons such effort is being made to isolate the kSZ contribution from primary fluctuations is to better understand the epoch of reionization. Gas density and velocity fluctuations during reionization had not yet formed nonlinear structure on galaxy and cluster scales, but the patchiness of the reionization process itself created order unity fluctuations in the electron density, and this is predicted to account for half or more of the total kSZ power. A prerequisite step for studying the EOR kSZ signal is to determine better the post-reionization part and subtract it from the total of the two components. Several attempts have been made to translate the observational kSZ limit into a constraint on the timing and duration of reionization \citep{2012ApJ...756...65Z,2012MNRAS.422.1403M,2013ApJ...776...83B}, although self-regulation of low-mass galaxies \citep{2013ApJ...769...93P} and the highly uncertain dependence on the mean free path of the percolation phase of reionization \citep{2012ApJ...747..126A}  creates a degeneracy in that parametrization. How to constrain the epoch of reionization with the kSZ signal is only partly understood and is an active area of investigation.  

The late-time temperature power spectrum is also the primary statistic that describes the overall scale dependence of fluctuations in the intergalactic medium, and understanding how it depends on baryonic processes in the post-reionization universe is of fundamental importance. 

In this paper, we utilize the Illustris simulation \citep{2015A&C....13...12N,2013MNRAS.436.3031V} as a case study for feedback and associated baryonic effects on electron density fluctuations and hence the kSZ power spectrum. Illustris is a recent cosmological hydrodynamic simulation of structure formation with a reasonably large ($75~h^{-1}~\rm{Mpc}$) sample volume, which has the advantage of being publicly available  with sufficient resolution to probe the {\em multi-phase gas distribution within galaxies} and groups on cosmological scales that includes the important effects determining the distribution of electrons, such as feedback from star formation and AGNs. The dark matter particle number and initial gas mesh number is $(1820)^3$, sufficient to resolve the small-scale  impact  on the kSZ effect.   

The kSZ signal is a line of sight integral of $-\hat{\gamma}\cdot \sigma_T n_e \mathbf{v}_e$, where $n_e$ is the electron density, $\mathbf{v}_e$  its peculiar velocity, $-\hat{\gamma}$  is the photon direction (towards us), and $\sigma_T$ is the Thompson cross section. In addition to using Illustris to better understand how electron density fluctuations evolve over time and are biased with respect to the dark matter, we also investigate the contribution of the 4-point {\it connected component} on the power spectrum, a decidedly nonlinear effect.  The current $n_e \mathbf{v}_e $ can be decomposed into the gradient of a potential and the curl of a vector. In Fourier space,  the gradient term is along the direction of the wavenumber, whereas the curl piece is perpendicular to it. The gradient component gives the linear components from recombination familiar from primary CMB anisotropies, but the kSZ power spectrum is primarily sourced by fluctuations in the curl component, $\mathbf{q_\perp}=[\mathbf{v}_e(1+\delta_e)]_\perp$. For full ionization, this is the transverse momentum field of the gas. The vorticity of the gas becomes nonzero when shocks occur, but is sub-dominant in its influence. Since our focus here is on the kSZ power spectrum, we want to study  the power spectrum of $\mathbf{q_\perp}=[\mathbf{v}_e\delta_e]_\perp$ in order to model the signal. Schematically this power spectrum is a four-point function of two densities and velocities, which can be decomposed into an unconnected part, involving the sum of products of two-point functions and a connected 4-point function, $\left<\delta v\delta v\right>_{c}$: $\left<qq\right>=\left<\delta\delta\right>\left<vv\right>+2\left<\delta v\right>^2+\left<\delta v\delta v\right>_{c}$. The connected  term arises when the density or velocity field becomes non-Gaussian.

Early analytic work focused on the linear regime calculation in which the density and velocity fields are Gaussian-distributed and the connected term does not exist \citep{1986ApJ...306L..51O, 1987ApJ...322..597V,1995ApJ...439..503D,1998PhRvD..58d3001J}. Later, calculations for the nonlinear density field were done by replacing the linear density power spectrum in the expression for the unconnected part, but the connected term was ignored \citep{2000ApJ...529...12H,2002PhRvL..88u1301M,2004MNRAS.347.1224Z,2012ApJ...756...15S}. Recently, a few studies reported that the connected term can add a significant amount of power in the cases where the density field becomes highly non-Gaussian. \citet{2016ApJ...818...37P}(hereafter P16) showed that nonlinear growth of structure can introduce  $\sim 30\%$ extra momentum power at $k\sim1~\rm{Mpc}^{-1}$, leading to $\sim10\%$ more post-reionization kSZ signal. For incomplete ionization during the reionization epoch, the ionization fraction is also associated with correlated processes distinct from the gas density, which should be described as a 6-point function and is almost impossible to estimate except through simulation.
\citet{2016ApJ...824..118A} measured a $\sim 30\%$ excess kSZ signal from the patchy ionized density field in their epoch of reionization simulation. It is reasonable to expect the baryonic physics would add to the non-Gaussianity in the plasma density field. We aim to confirm that hypothesis in this work. Maps made of the kSZ effect with concentrated regions of electron density in groups and clusters and with velocities self-consistently included have for a long time included aspects of the nonzero connected 4-point pieces. 


The paper is organized as follows. In Section 2, we express the kSZ signal in terms of transverse momentum power spectrum and describe how the unconnected terms of transverse momentum power spectrum are related to the density and velocity power spectra. In Section 3, we describe the Illustris simulation, which we will mainly use for our analyses. In Section 4, we quantify how the baryonic physics affect the plasma density power spectra, and contrast it with the momentum power for dark matter. We then measure the contribution of the connected term in the transverse momentum power spectrum from the simulation. We also report the post-reionization kSZ signal from the simulation along with values reported by other recent works. In section 5, we summarize and discuss our results.

\section{The KSZ signal from the Post-reionization Era}

\subsection{The Angular Power Spectrum of the kSZ Effect}

The relative change in the CMB temperature induced by the kSZ effect in the direction of the line-of-sight unit vector $\hat{\gamma}$ is
\bea \label{eq:kSZ}
\frac{\Delta T (\hat\gamma)}{T} = -\int d\tau e^{-\tau} \frac{\hat\gamma \cdot \mathbf{v}_e}{c},
\eea
where $\mathbf{v}_e$ is the peculiar velocity field of the electrons, $\tau$ is the (inhomogeneous) Thompson optical depth between $z = 0$ and the scatterer, where $d\tau=c ~n_{e} \sigma_T dt$. In the post-reionization universe, the IGM is nearly fully ionized and $n_e$ closely follows the underlying gas density field. We define the free electron number density contrast by 
\bea \label{eq:delta}
\delta_{e} = \frac{n_e}{\bar{n}_e} - 1, 
\eea
in terms of the global average of the electron density $\bar{n}_e$. Defining the {\it specific momentum} (henceforth referred to as ``momentum") $\mathbf{q}\equiv \mathbf{v}_e[1+\delta_e]$ allows us to rewrite Equation~(\ref{eq:kSZ}) as 
\bea \label{eq:kSZ2}
\frac{\Delta T (\hat\gamma)}{T} = - \frac{\sigma_T \bar{n}_{e,a0}}{c} \int \frac{ds}{a^2} e^{- \tau} f_e \mathbf{q} \cdot {\hat\gamma},
\eea
where $s(z)$ is the comoving distance to the scatterer and  $\bar{n}_{e,a0}\equiv \bar{n}_{e,a}(z=0)$ is the current redshift 0 value of the mean number density of free plus bound electrons, $\bar{n}_{e,a}(z)\equiv (1- Y_P/2)\bar\rho_b(z)/m_p$. The mean baryon number density is $\rho_b(z)/m_p$, where  $\bar\rho_b(z)=[3H_0^2\Omega_b (1+z)^3]/[8\pi G]$ is the global baryonic mass-density of the universe and $m_p$ is the proton mass. We set the primordial helium abundance to $Y_P=0.24$; the small metal contamination for these average numbers is ignored. We define the fraction of free electrons to be $f_e(z)\equiv \bar{n}_e/\bar{n}_{e,a}$. Thus if Helium is fully ionizes as well as hydrogen  $f_e$ would be unity. 

As mentioned in the introduction, the Fourier transform of the momentum, $\tilde{\mathbf{q}}\equiv \int d^3 \mathbf{x} ~e^{i \mathbf{k} \cdot \mathbf{x}} \mathbf{q}(\mathbf{x})$, can be separated into the transverse component, $\tilde{\mathbf{q}}_\perp \equiv \tilde{\mathbf{q}} - \mathbf{k}[\tilde{\mathbf{q}} \cdot \mathbf{k}]$, which is perpendicular to the wavevector, and the longitudinal component, $\tilde{\mathbf{q}}_\parallel \equiv \mathbf{k}[\tilde{\mathbf{q}} \cdot \mathbf{k}]$, which is parallel to the wavevector. In the kSZ angular power spectrum of $\Delta T /T$, $\tilde{\mathbf{q}}_\parallel$ gives a negligibly small contribution compared to that from $\tilde{\mathbf{q}}_\perp$  \citep{1987ApJ...322..597V,2016ApJ...824..118A,2016ApJ...818...37P}, though it dominates the $\Delta T /T$  primary anisotropies during recombination and at low multipoles $\ell < 10$ through reionization.  Thus we can write the  CMB angular power spectrum in terms of the transverse component only:
\bea \label{eq:kSZ_Cl}
\mathcal{C}_\ell = \frac{1}{2} \left[\frac{\sigma_T  \bar{n}_{e,a0}}{c} \right]^2 \int \frac{ds}{s^2 a^4} e^{-2\tau} f_e^2 P_{q_\perp} \left( k=\frac{l}{s}\right)~~~
\eea
\citep{1987ApJ...322..597V}. Thus  $f_e(z)$ and $P_{q_\perp}(k,z)$ are the key quantities that determine the shape and amplitude of the kSZ power spectrum for a fixed background cosmology.

\subsection{Power Spectrum of the Transverse Momentum Field}

Here, we shall summarize the derivation of $P_{q_\perp}$ in terms of the density and velocity power spectra. In Fourier space 
\bea
\tilde{\mathbf{q}} = \tilde{\mathbf{v}} + \int \frac{d^3k^\prime}{(2\pi)^3} \tilde{\delta}_e (\mathbf{k}-\mathbf{k^\prime}) \tilde{\mathbf{v}} (\mathbf{k^\prime}),
\eea
where $\tilde{\mathbf{q}} = \int d^3 \mathbf{x} ~e^{i \mathbf{k} \cdot \mathbf{x}} \mathbf{q}(\mathbf{x})$  . 

\begin{figure*}
  \begin{center} 
    \includegraphics[scale=0.48]{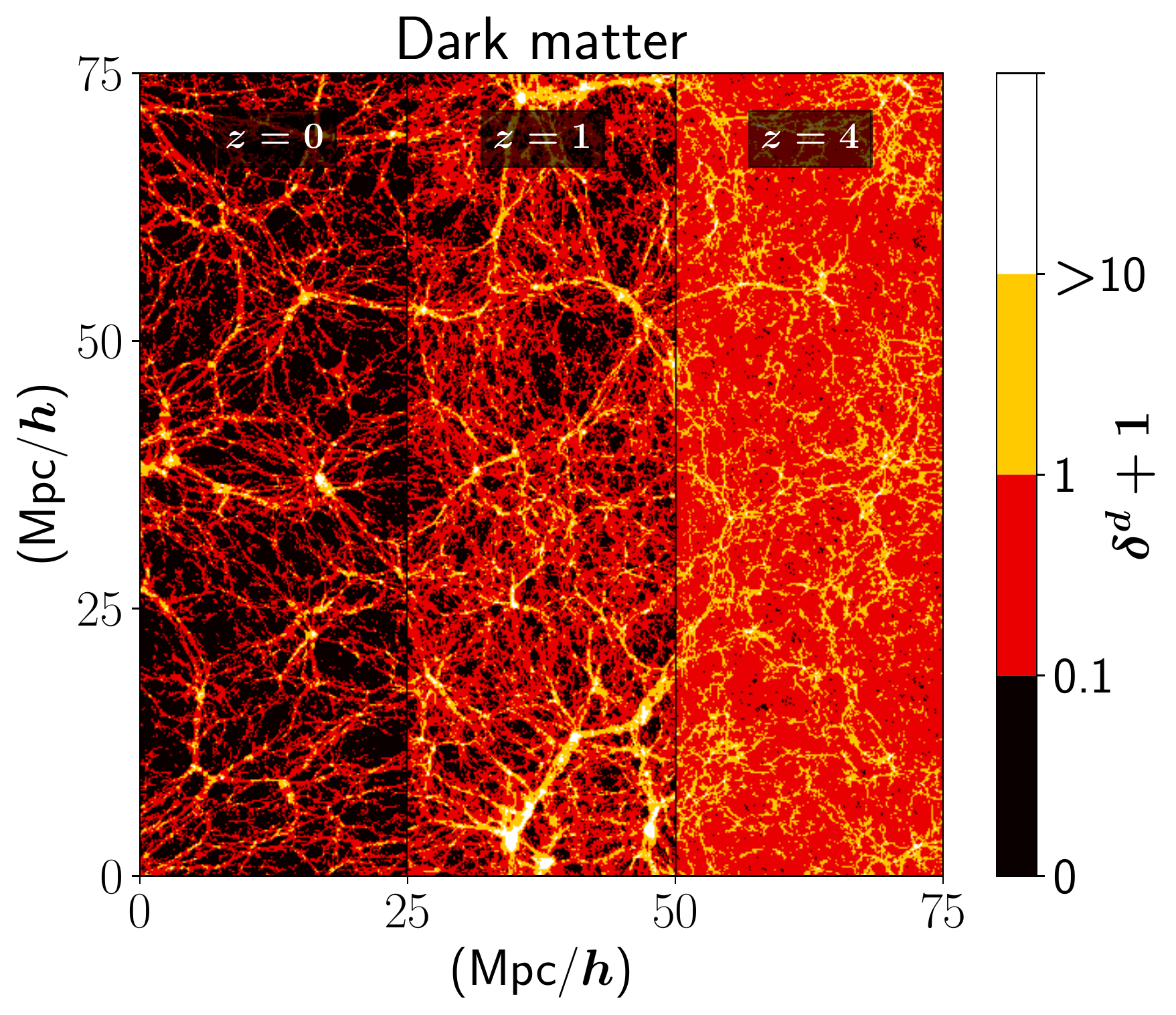}
    \includegraphics[scale=0.44]{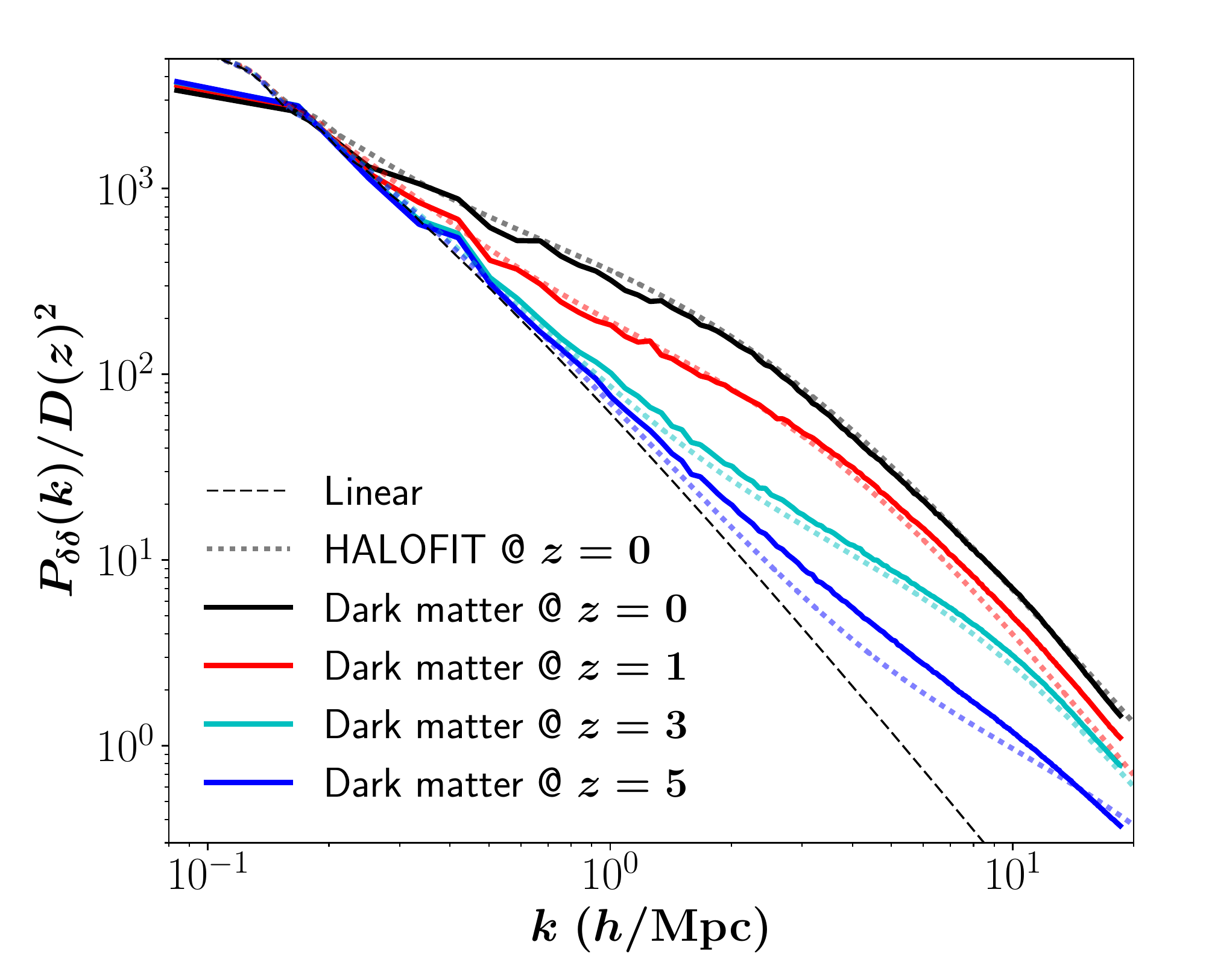}
    \includegraphics[scale=0.48]{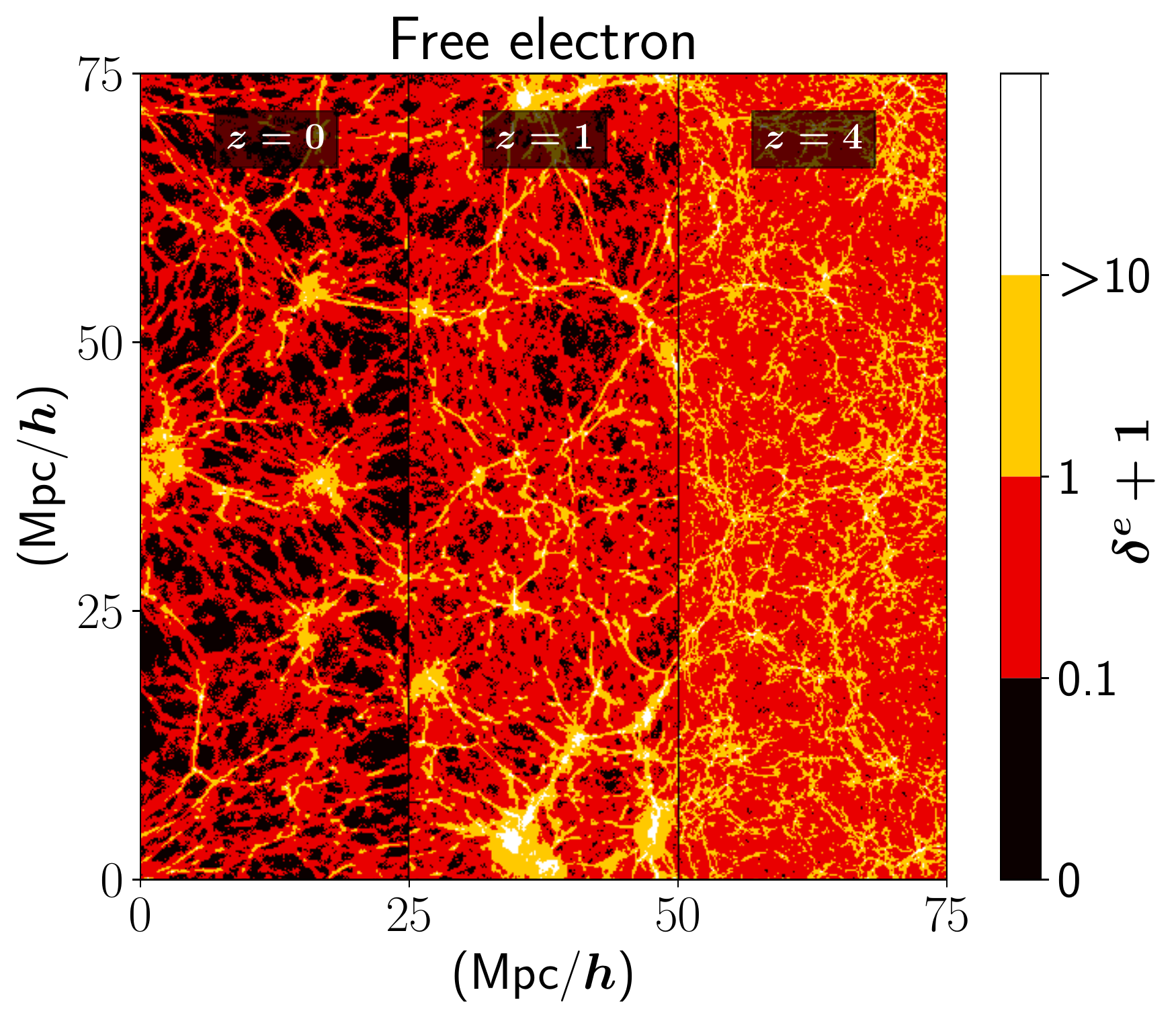}
    \includegraphics[scale=0.44]{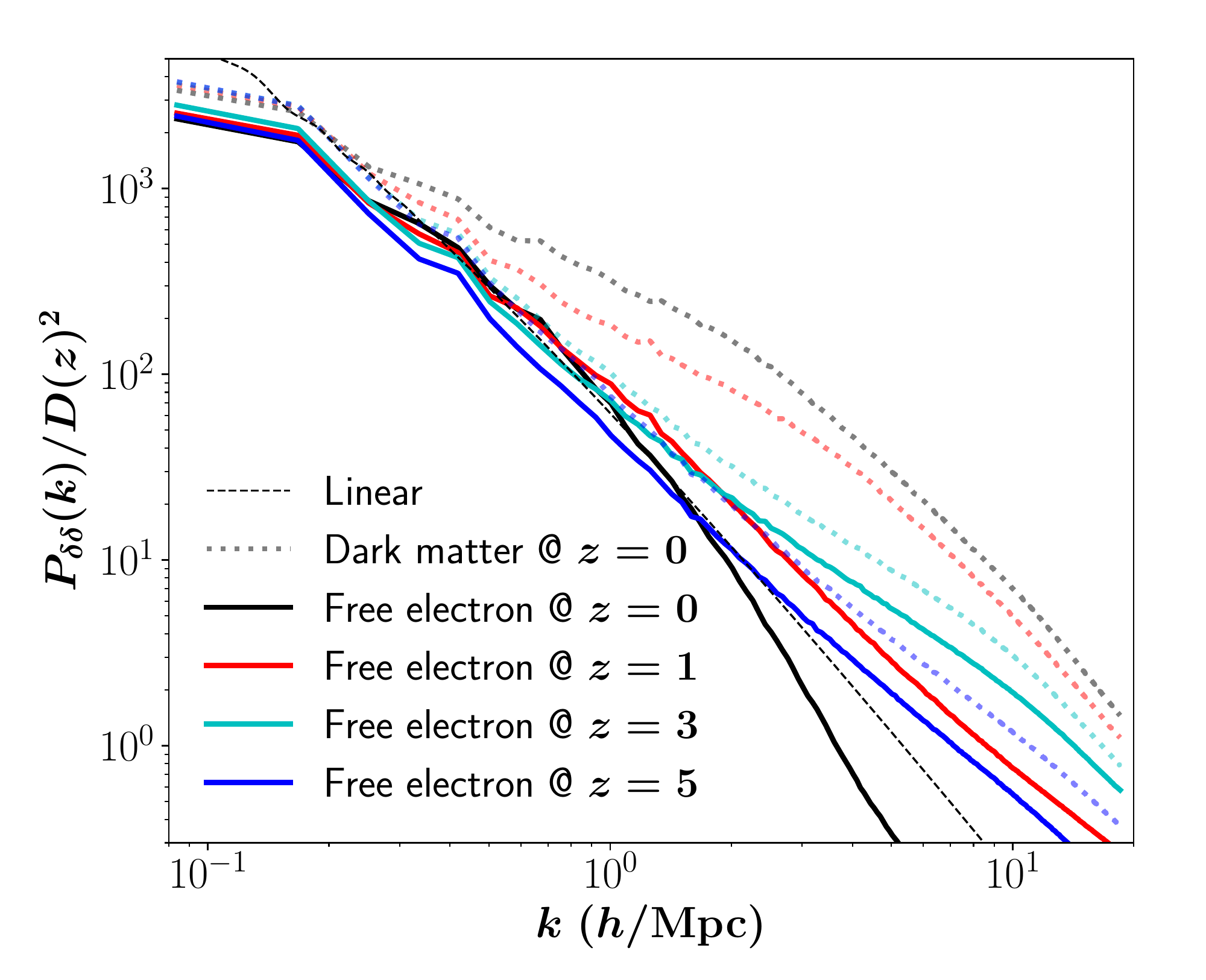}
  \caption{ (Upper left) Density contrast map of a one-cell thick slice from a $512^3$ dark matter density mesh for $z=0,1$ and 4 from left to right. (Lower left) A similar density contrast map for a free electron density mesh. (Upper right) The density power spectrum of the Illustris dark matter density field is shown divided by the growth factor squared as solid lines for $z=0$ (black), 1 (red), 3 (cyan), and 5 (blue). For comparison, the result from the HALOFIT model is plotted as dotted lines for the same redshifts with same colors. The linear theory expectations are the dashed lines. (Lower right) Similar to the upper right panel, but using solid lines for the Illustris free electron density field and dotted lines for the Illustris dark matter density field.}
  \label{fig:Den_PS}
  \end{center}
\end{figure*}

The first term can have non-zero curl only in the nonlinear regime, but its contribution is below a percent level \citep[][P16]{2002PhRvL..88u1301M}. The transverse power spectrum from the second order term, $\int \delta v$, gives
\bea
P_{q_\perp}(k) &=& \int \frac{d^3 k^\prime}{(2\pi)^3} \int \frac{d^3 k^{\prime\prime}}{(2\pi)^3} [\hat{\mathbf{k^\prime}} \cdot \hat{\mathbf{k^{\prime\prime}}} - (\hat{\mathbf{k}} \cdot \hat{\mathbf{k^\prime}})(\hat{\mathbf{k}} \cdot \hat{\mathbf{k^{\prime\prime}}})] \nonumber \\
&&P_{\delta\delta v v} (\mathbf{k} - \mathbf{k^\prime}, -\mathbf{k} - \mathbf{k^{\prime\prime}}, \mathbf{k^\prime}, \mathbf{k^{\prime\prime}}),
\eea
where $P_{\delta \delta v v}$ is given by
\bea
&&(2\pi)^3 P_{\delta \delta v v} (\mathbf{k_1},\mathbf{k_2},\mathbf{k_3},\mathbf{k_4}) \delta_D (\mathbf{k_1}+\mathbf{k_2}+\mathbf{k_3}+\mathbf{k_4}) \nonumber \\
&\equiv& \left< \tilde{\delta}(\mathbf{k_1})  \tilde{\delta}(\mathbf{k_2})  \tilde{v}(\mathbf{k_3})  \tilde{v}(\mathbf{k_4})\right>.
\eea

In the linear regime, all of the four fields in the bracket, $\left<\delta\delta v v\right>$, are Gaussian if the initial conditions are. This allows us to expand the whole bracket into brackets with two fields using  Wick's theorem: $\left<ABCD\right>_{uc}=\left<AB\right>\left<CD\right>+\left<AC\right>\left<BD\right>+\left<AD\right>\left<BC\right>$. In the post-reionization regime, however, nonlinearity creates non-Gaussianity in the density and velocity fields, which gives a rise to the connected term, $\left< ABCD\right>_{c}$.

The unconnected part is given by
\bea \label{eq:MF}
&&P_{q_{\perp},uc}(k,z) = 
\int \frac{d^3k^\prime}{(2\pi)^3} (1-{\mu^\prime}^2) 
\left[
P_{\delta\delta}(|\mathbf{k}-\mathbf{k^\prime}|)
P_{vv} (k^\prime) 
\right.
\nonumber \\ 
&&\left. \qquad \qquad \qquad \qquad 
-\frac{k^\prime}{|\mathbf{k}-\mathbf{k^\prime}|}
P_{\delta v}(|\mathbf{k}-\mathbf{k^\prime}|)
P_{\delta v} (k^\prime) 
\right],
\eea 
where $\mu^\prime \equiv \hat{\mathbf{k}}\cdot\hat{\mathbf{k}^\prime}$. Instead of the above expression, the following approximate expression is often used since it is highly accurate and convenient to evaluate \citep[][P16]{2000ApJ...529...12H}.
\bea \label{eq:Pq_unconnected}
P_{q_{\perp},uc}(k,z) &=& \int \frac{d^3k^\prime}{(2\pi)^3} P^e_{\delta\delta} (|\mathbf{k} - \mathbf{k^\prime}|,z) P_{vv}(k^\prime,z)\nonumber\\
&&\qquad \qquad\times
\frac{k(k - 2k^\prime\mu^\prime)(1-{\mu^\prime}^2)}{k^2 + {k^\prime}^2-2kk^\prime\mu^\prime},
\eea
For higher wavenumbers  $k\gtrsim 0.1~h~\rm{Mpc}^{-1}$  the kSZ signal is still quite relevant, and most of the contribution in the above integral comes from the coupling of $P_{vv}(k^\prime)$ at  $k^\prime\lesssim 0.1~h~\rm{Mpc}^{-1}$ with $P_{\delta\delta} (|\mathbf{k} - \mathbf{k^\prime}|)\approx P_{\delta\delta}(k)$.  While linear calculations are accurate enough for $P_{vv}$, we need a more accurate $P_{\delta\delta}$  to account for nonlinear effects at high $k$'s. 

At $z\lesssim3$ and $k>1~h~\rm{Mpc}^{-1}$, $P_{\delta\delta}$ is strongly affected by both the growth of nonlinear structure, which can differ for gas cf. dark matter, and the baryonic physics operating. We characterize how  the electron density power $P^e_{\delta\delta}$ differs from the dark matter power $P^d_{\delta\delta}$ by a scale and redshift dependent bias, defined here by 
\bea
b_e(k,z)\equiv \sqrt{\frac{P^e_{\delta\delta}(k,z)}{P^d_{\delta\delta}(k,z)}}. 
\eea
To see the combined effects in both $f_e$ and $P_{q_\perp}$, \citet{2012ApJ...756...15S} (hereafter SRN12) introduced a window function,
\bea \label{eq:W}
W_e^2(k,z)\equiv f_e^2(z) b_e^2(k,z),
\eea
which {\bf we} adopt in this work to describe the baryonic effects.





We note that $P_{q_{\perp},uc}$ does not fully account for $P_{q_{\perp}}$ in the nonlinear regime due to the rise of the connected term, $P_{q_{\perp},c}$. P16 confirmed that it is present and important to include for the pure dark matter momentum field. We find the connected term is still important for the electrons, but differs  in detail because of the relative bias.

\section{Simulation}

For our analysis of the power, and its deconstruction into connected and unconnected parts, we use the publicly available results from Illustris, a hydrodynamic simulation of galaxy formation from cosmological initial conditions in a $(75~h^{-1}~\rm{Mpc})^3$ comoving volume periodic box. Its hydrodynamics and gravitational dynamics are driven by the moving mesh code, AREPO \citep{2010MNRAS.401..791S}. It uses sub-grid prescriptions for chemical cooling of the gas, subsequent star and black-hole formation, and the feedback of stars and black-holes on the surrounding medium. The details of the sub-grid models are described in \citet{2013MNRAS.436.3031V} and \citet{2014MNRAS.438.1985T}. The simulation starts with $1820^3$  dark matter particles, each having $6.26\times 10^6~M_\odot$. The  number of resolution elements for the gas is the same, with the initial mass of $1.26\times 10^6~M_\odot$. This setup allows halos to be resolved down to $10^8~M_\sun$ in mass,  and  power spectra to be reliable up to $k\sim 10~h~\rm{Mpc}^{-1}$ in wavenumber. The cosmology is based on a tilted $\Lambda$CDM model with $\Omega_m = 0.2726$, $h=0.704$, $\Omega_b = 0.0456$, $\sigma_8 = 0.809$, and $n_s=0.963$. 

Illustris aims to provide a realistic sample of the universe by matching galactic properties with observation. Halos in Illustris are identified by a standard friends-of-friends (FoF) algorithm with linking length $b=0.2$. Then the SUBFIND algorithm of \citet{2001MNRAS.328..726S} is used to identify subhalos, which are regarded as galaxies. Illustris is known to reproduce the stellar mass function and the luminosity function of galaxies reasonably well  and results in a realistic morphological composition of galaxies. The key results from Illustris on galactic properties are discussed in \citet{2014Natur.509..177V} and \citet{2014MNRAS.445..175G}.

The snapshots of Illustris are readily available on the web. We use particle snapshots for $z=0,0.5,1,2,3,4,$ and $5$ in this work. The snapshots contains masses, locations and velocities of dark and gas particles, and free electron fractions, required for computing the  transverse momentum power spectra of the species present. 

\section{Results}

\subsection{Density Power Spectrum of Dark Matter and Free Electron}   \label{Sec:DPS}

We show the power spectra of dark matter ($P^d_{\delta\delta}$) and free electron ($P^e_{\delta\delta}$) density fields in the right panels of Figure~\ref{fig:Den_PS} for $z=0,~1,~3,$ and 5.  Nonlinearity results in the familiar result of $P^d_{\delta\delta}$ growing faster than the linear theory prediction. At $k>1~h~\rm{Mpc}^{-1}$, $P^d_{\delta\delta}$ is about an order of magnitude larger than the linear prediction at $z=0$. As is well known from previous works, the HALOFIT model of \citet{2003MNRAS.341.1311S} does a good job in modeling this enhancement. The evolution of $P^e_{\delta\delta}$, however, is more complicated, showing the effects of dynamical  nonlinear growth at $z\ge3$, but complicated by separation of gas from baryons by shocking and feedback, combining to give a growth slower than linear after $z=3$. How this difference between $P^d_{\delta\delta}$ and $P^e_{\delta\delta}$ arises due to the effects of baryonic physics of plasma is what we aim to probe.

A list of relevant baryonic physics that will impact the free electrons relative to the dark matter are: 
\begin{itemize}
\item[1)] Collapse of cooled gas
\item[2)] Star-formation converting gas into stars and blackholes
\item[3)] AGN feedback blowing gas out of potential wells
\item[4)] Self-shielding of cold neutral gas in galaxies against ionizing radiation
\item[5)] He II reionization at $z\sim3$
\end{itemize}
For example, the density map of free electron in comparison to that of dark matter (See the left panels of Figure~\ref{fig:Den_PS}) shows the smoothing of the electron density field {\bf at} $\boldsymbol{z\le 1}$ at certain scales that we can attribute to the effect of AGN feedback. The spread extends over larger scales than in some other gasdynamical calculations, a consequence of the strong feedback in Illustris. This explains the deficit in $P^e_{\delta\delta}$ compared to $P^d_{\delta\delta}$ at high $k$ and low redshift. To further analyze the differences, we now explore the window function $W_e$ (Eq.~\ref{eq:W}).


\newcommand*{\myfont}{\fontfamily{phv}\selectfont}
\begin{figure}
  \label{fig:Wb}
  \begin{center}
  \includegraphics[scale=0.5]{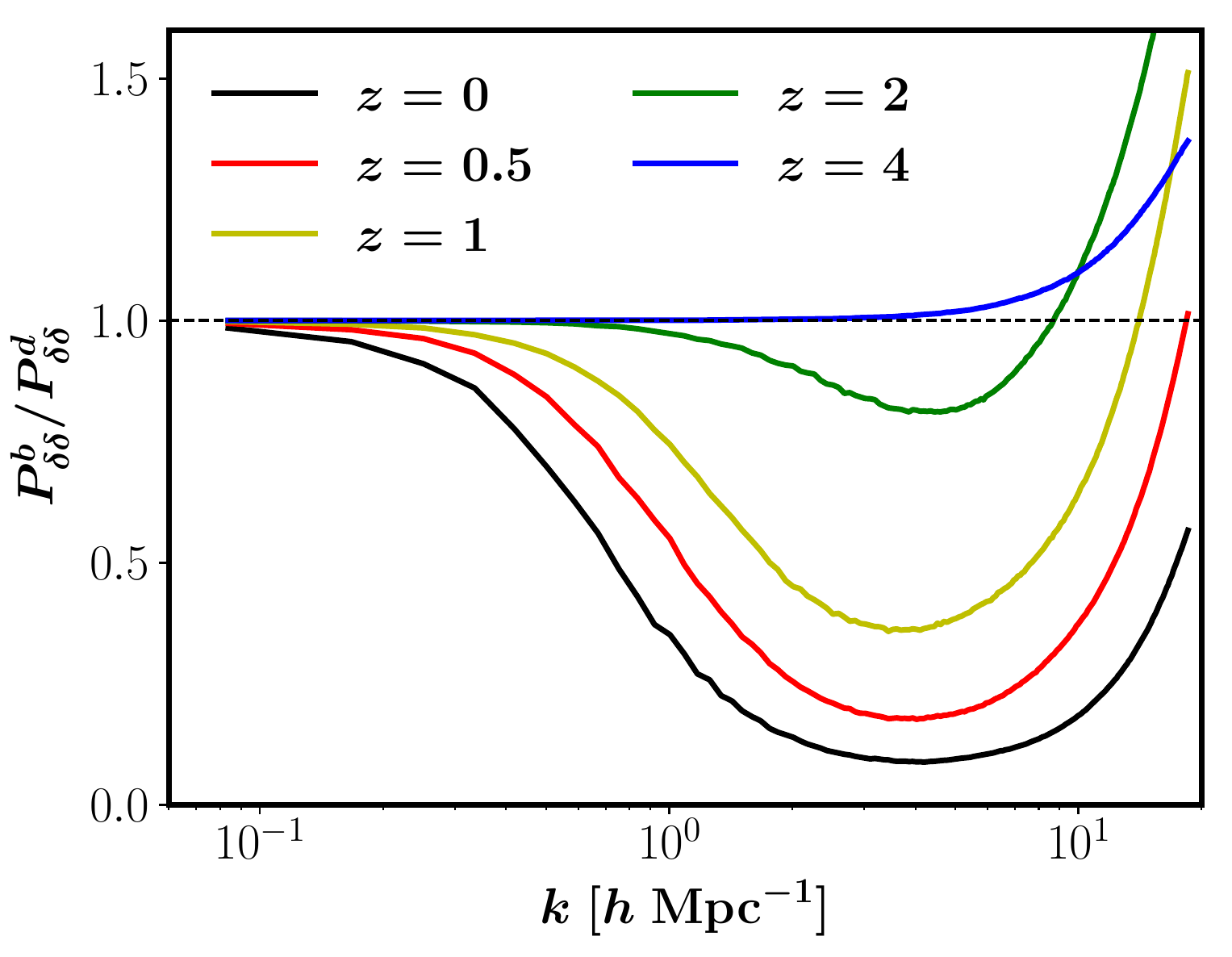}
  \caption{Scale-dependent bias of baryon fluctuations with respect to dark matter, $b_b^2(k)\equiv P_{\delta\delta}^b/P^d_{\delta\delta}$, at $z=0$ (black), 0.5 (red), 1 (yellow), 2 (green), and 4 (blue). }
  \end{center}
\end{figure}

\subsection{Window Functions of Baryon, Gas, and Free Electrons}

\begin{figure*}
  \begin{center}
    \includegraphics[scale=0.5]{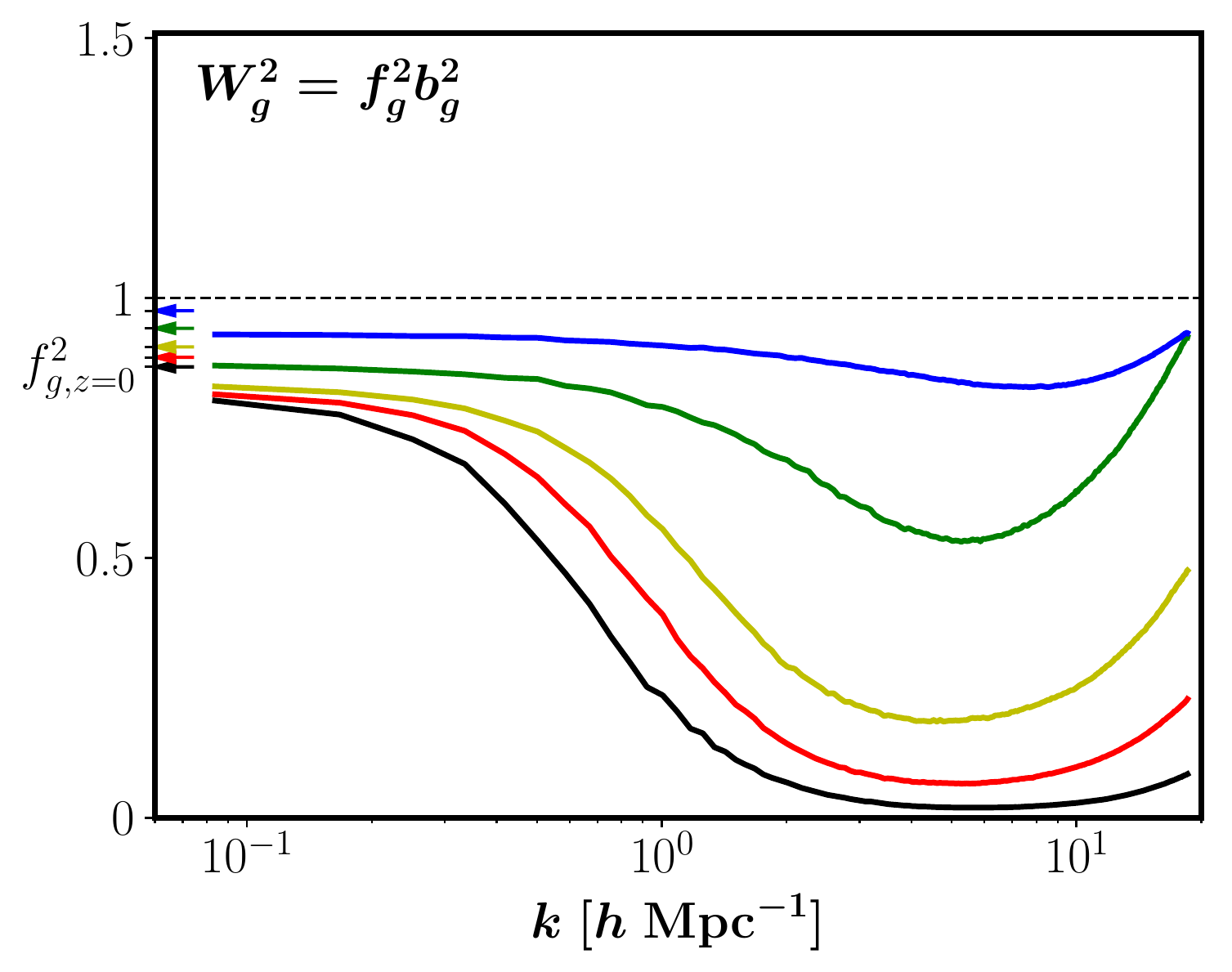}
    \includegraphics[scale=0.5]{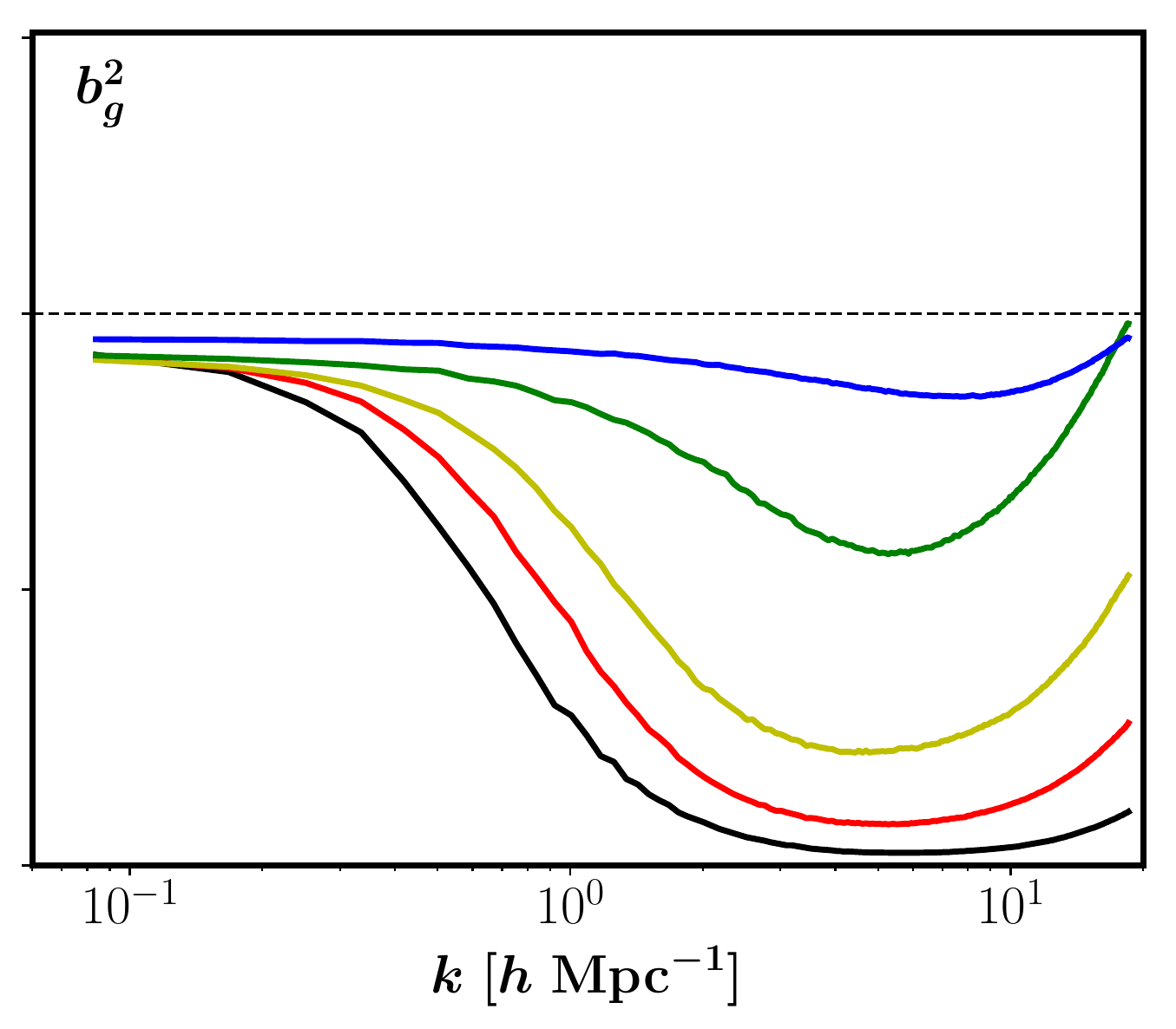}
    \includegraphics[scale=0.5]{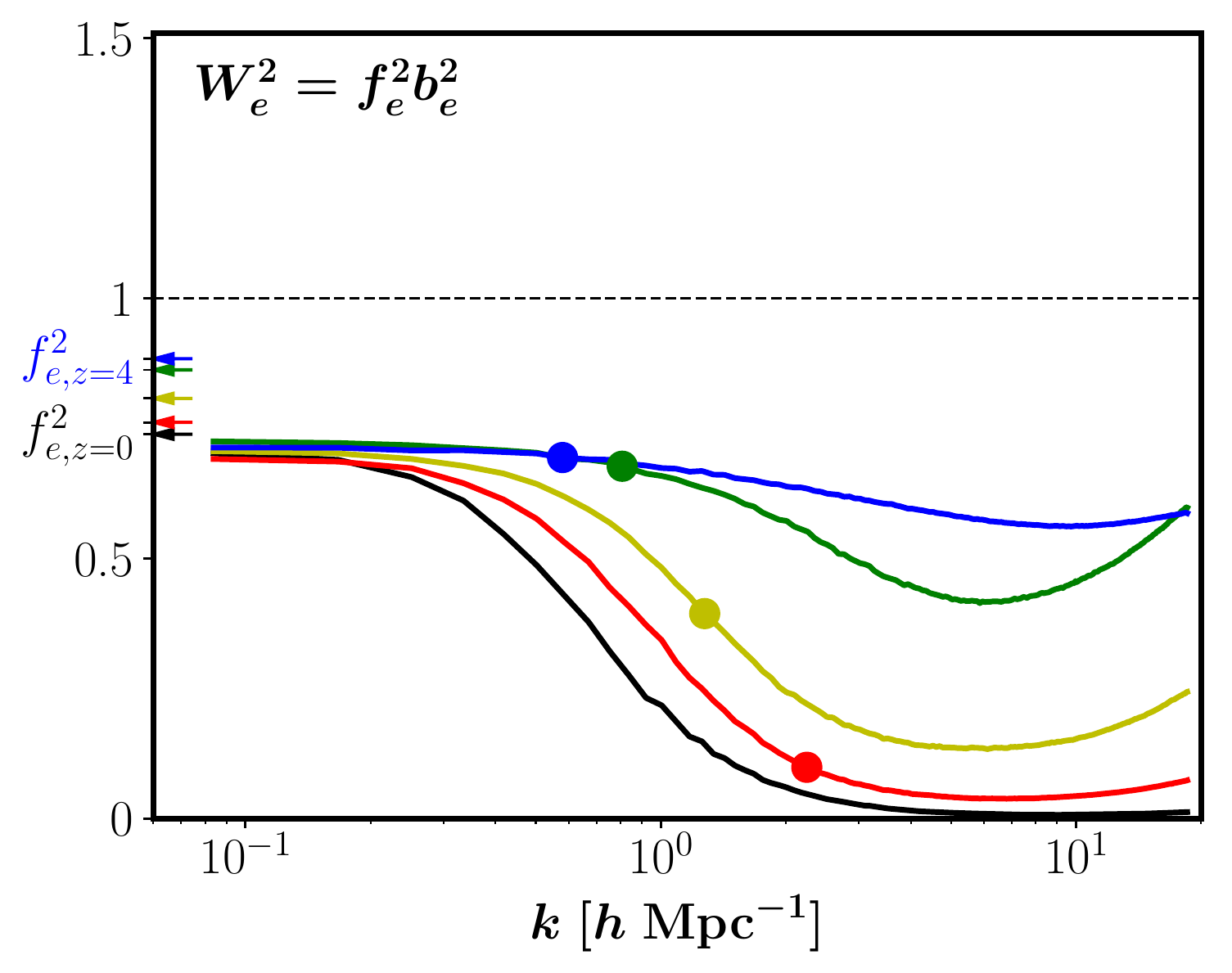}
    \includegraphics[scale=0.5]{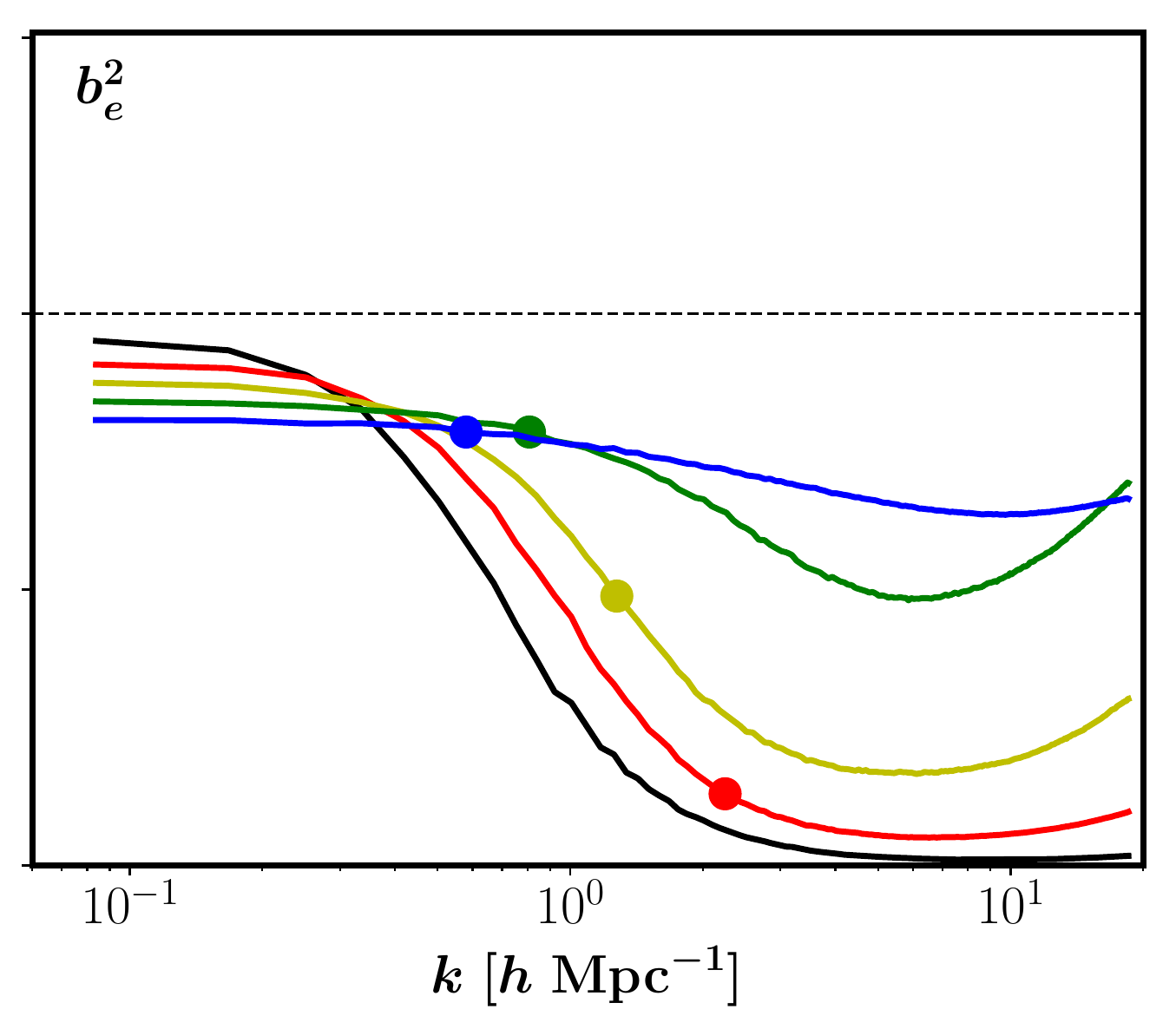}
  \caption{Window functions (left) and bias (right) for gas density (top) and electron density (bottom). Indicated with arrows on $W_g^2$ and $W_e^2$ are the values of $f_g^2$ and $f_e^2$, respectively. The colors of the arrows represent redshift with the same color scheme as in the curves, as shown in Figure 2. Filled circles on the curves in the bottom panels denote the wavenumbers that contribute to the $\ell=3000$ signal (using $k =\ell/s(z)$, where $s(z)$ is the comoving distance) at given redshifts. Color scheme is same as for Figure 2:  $z=0$ (black), 0.5 (red), 1 (yellow), 2 (green), and 4 (blue). $f_g^2$ and $f_e^2$ are shown as extra y-ticks between 0.5 and 1 with arrows attached in the right-hand-side for $W_g^2$ and $W_e^2$, respectively. The colors of the arrows represents redshift with the same color scheme as in the curves.
  $b_g^2$ and $b_e^2$ are shown in the middle left and bottom left panels, respectively. Filled circles on the curves in bottom panels denote the wavenumbers that contribute to the $\ell=3000$ signal ($\ell/s(z)$) at given redshifts.}
     \label{fig:W}
   \end{center}
\end{figure*}

The window function for the free electrons relative to dark matter $W_e$ encodes the effect of  all of the baryonic physics relevant to the intergalactic plasma. To help deconstruct its shape  we explore  equivalent window functions for baryons ($W_b$) and gas ($W_g$) as well, defined similarly to $W_e$, except that we use baryon  and gas densities are used instead of electron number density in Equation~(\ref{eq:delta}). The baryon density field includes all the mass in stars, AGNs and neutral gas as well as gas and we can thus use it to exclude the effect of converting gas into those elements that do not give rise to the kSZ effect. The gas density field does include neutral gas in galaxies, but not stars or AGNs, which allows us to decompose the effect of having multiphase gas. We show $W_b,~W_g,$ and $W_e$ in Figure~\ref{fig:W} for $z=0,~0.5,~1,~2,$ and 4. 

 \begin{figure}
  \begin{center}
    \includegraphics[scale=0.5]{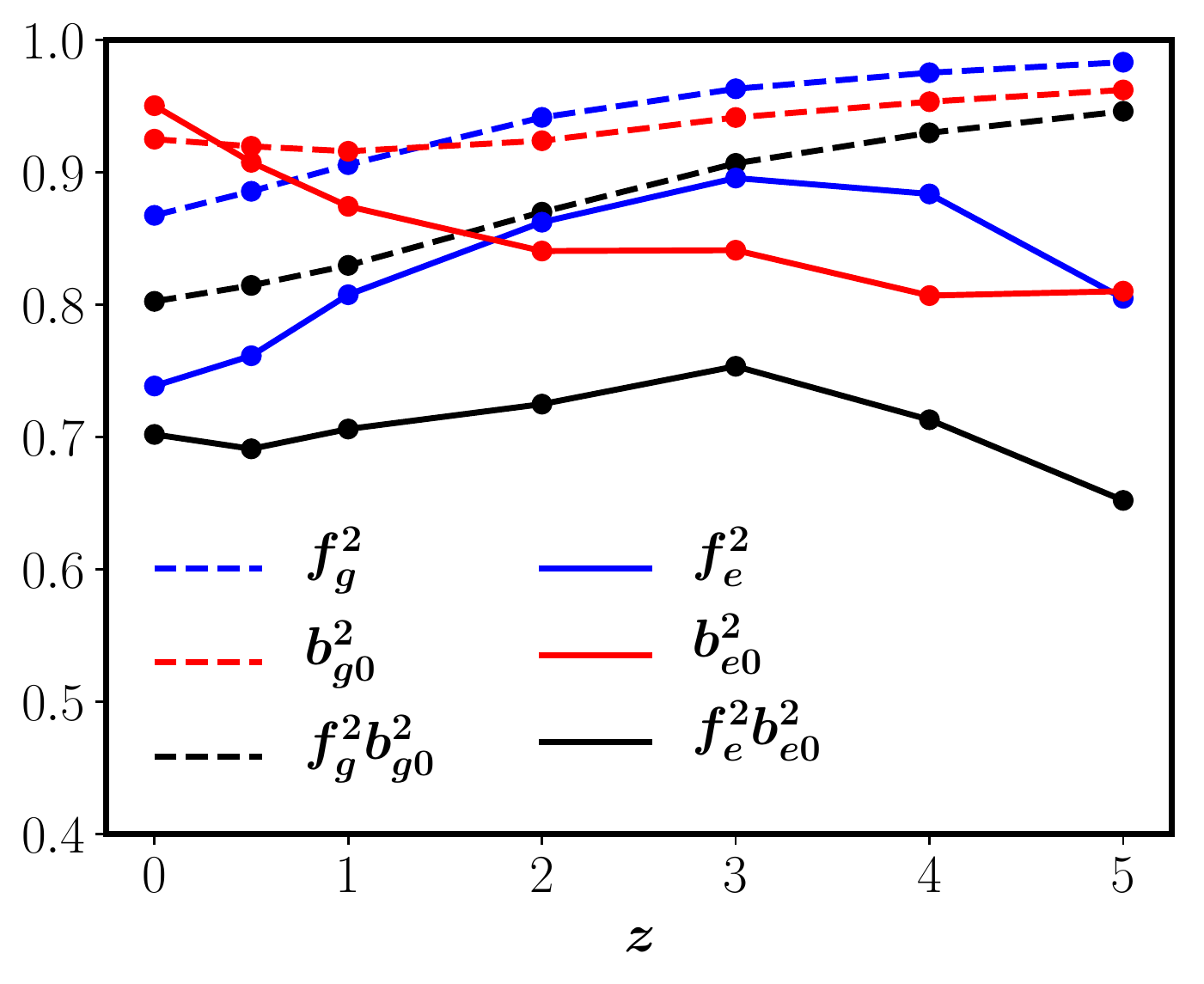}
  \caption{Abundances (blue), large-scale bias (red), and total suppression (black), as a function of redshift for electrons (solid) and gas (black). The large-scale suppression factor for the electron density fluctuations, with respect to the case with no stars and fully ionized gas, is quite constant, varying between $\sim$ 0.65 and 0.75.}
  \label{fig:fb}
  \end{center}
\end{figure}

In $W_b$ an up-turn appears at $k\sim 5~h~\rm{Mpc}^{-1}$ at all redshifts shown, a trend easily understood by the collapse of gas on small scales following gas cooling in over-dense regions. At $z\le2$, the feedback of star-formation and AGNs heats the gas and makes it expand over time, damping $W_b$ at $k\gtrsim 0.3~h~\rm{Mpc}^{-1}$. The combination of these two effects creates a dipper-like shape in $W_b^2$ at $z\le1$.

The comparison of $W_g$ to $W_b$ demonstrates the effect of star-formation. As gas is converted into stars in density peaks, gas density power is strongly suppressed at $k\gtrsim1~h~\rm{Mpc}^{-1}$. As a result, the up-turn at high-$k$ is much weaker in $W_g$. Additionally, star-formation suppresses $W_g$ at all wavenumbers as can be seen from the low-$k$ limit of $W^2_g$ being less than one. For the case in which a fraction  $(1-f_g)$ of gas was of converted into stars homogeneously, $W_g^2$ would be suppressed by a factor of $f_g^2$. In Illustris we find  $W^2_g$ drops below  $f_g^2$ at low $k$ because star-formation drains gas preferentially in over-densities and create an anti-bias in the gas density field. The bias of the gas density $b_g$ shown in the middle right panel of Figure~\ref{fig:W} clearly shows this anti-biasing effect in its low-$k$ limit. To extract the anti-bias, we define $b_{g0}$ as $W_g(k\rightarrow 0)$. In Figure~\ref{fig:fb} we plot $b^2_{g0}$, $f^2_g$, $f^2_g b^2_{g0}$ and their counterparts for electron density. Due to ongoing star-formation, $f^2_g$ monotonically decreases from 0.92 to 0.8 between $z=5$ and $0$. $b_g^2$ starts from 0.96 at $z=0$ and decreases until $z=1$ down to 0.92 and stays without much change until $z=0$. The combination of these two term, $f^2_g b^2_{g0}$, gives a large-scale suppression effect that grows from $5\%$ to $20\%$ between $z=5$ and $0$.

 $W_e$ has additional suppression effects on top of what it inherits from $W_g$ because not all electrons that survived star-formation contribute to  the mean ionized  plasma: a significant fraction of them are trapped in the cold neutral gas in galaxies that remains self-shielded from extragalactic ionizing radiation or are locked in helium atoms at $z\gtrsim3$ where most heliums are only singly ionized even when hydrogen is ionized. Similarly to the comparison of $W_g$ to $W_b$, $W_e$ gets an all-wavenumber suppression plus more high-$k$ suppression compared to $W_g$, which makes $W_e<W_g$ at all scales. The filled circles on the curves for $W_e$ denote the wavenumber that contributes mostly to the $\ell=3000$  signal at give $z$, $\ell/s(z)$. Those circles show that the signal down to $z=2$ is mostly affected by the large-scale term $f^2_g b^2_{g0}$ and the $z<2$ signal is further suppressed by the AGN feedback.

\subsection{Susceptibilities of Gas and Free Electron to Dark Matter: Response Functions compared to Window Functions}

 We have  paid considerable attention to the  deconstruction of  the total kSZ power into some of its two point subcomponents, in particular targeting how the $\langle \delta \delta \rangle$ terms vary relative to each other for the electron density, gas density and dark matter through the power ratios $b^2_e = P^e_{\delta\delta}/P^d_{\delta\delta}$ and $b^2_g = P^g_{\delta\delta}/P^d_{\delta\delta}$, Fourier transforms of the position-position correlations $[\langle \delta_e \delta_e \rangle \langle \delta_d \delta_d\rangle^{-1} ](\mathbf{x}-\mathbf{x}^\prime)$ and $[\langle \delta_g \delta_g \rangle \langle \delta_d \delta_d\rangle^{-1} ](\mathbf{x}-\mathbf{x}^\prime)$. Of interest as well are statistically averaged values of $n_e$ and $\rho_g$ subject to the constraint of $\rho_d$. This is a nonlinear relationship which involves averaging over constrained probability functionals to 
 yield $\langle \delta_e (\mathbf{x}) \vert \delta_d \rangle$ and $\langle \delta_g (\mathbf{x}) \vert \delta_d \rangle$. However there is a ``linear response" limit in which, e.g., 
\bea 
\langle n_e (x) | \rho_d \rangle / \langle n_e\rangle &&= \int \chi_{ed} (x-x^\prime) \rho_d (x^\prime) d^3 x^\prime  /\langle \rho_d \rangle,  \\
\chi_{ed} (\mathbf{x}-\mathbf{x}^\prime) &&= [\langle \delta_e \delta_d \rangle \langle \delta_d \delta_d\rangle^{-1} ](\mathbf{x}-\mathbf{x}^\prime) \, \nonumber.
\eea 
The (suitably normalized) susceptibility of $n_e$ to $\rho_d$ is $\chi_{ed}$, with units of inverse volume; the name is chosen because there is a clear relation to the familiar  susceptibilities  that relate electric and magnetic polarizabilities to applied electromagnetic fields, ultimately leading to dielectric tensors. Another terminology for $\chi_{ed}$ is response function, for obvious reasons. 
 \begin{figure*}
  \begin{center}
    \includegraphics[scale=0.5,trim = 0 0 0 0, clip=true]{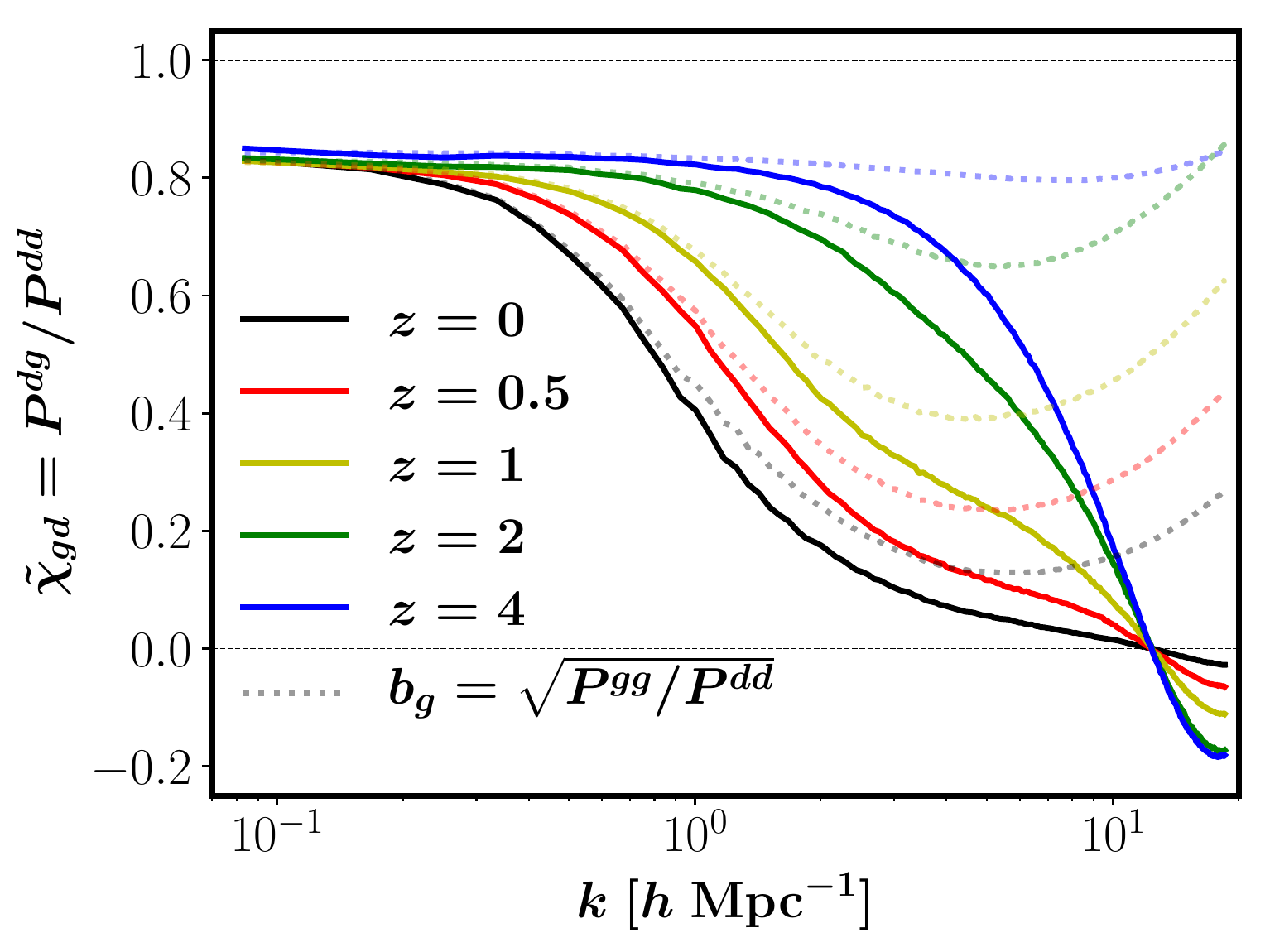}
    \includegraphics[scale=0.5,trim = 0 0 0 0, clip=true]{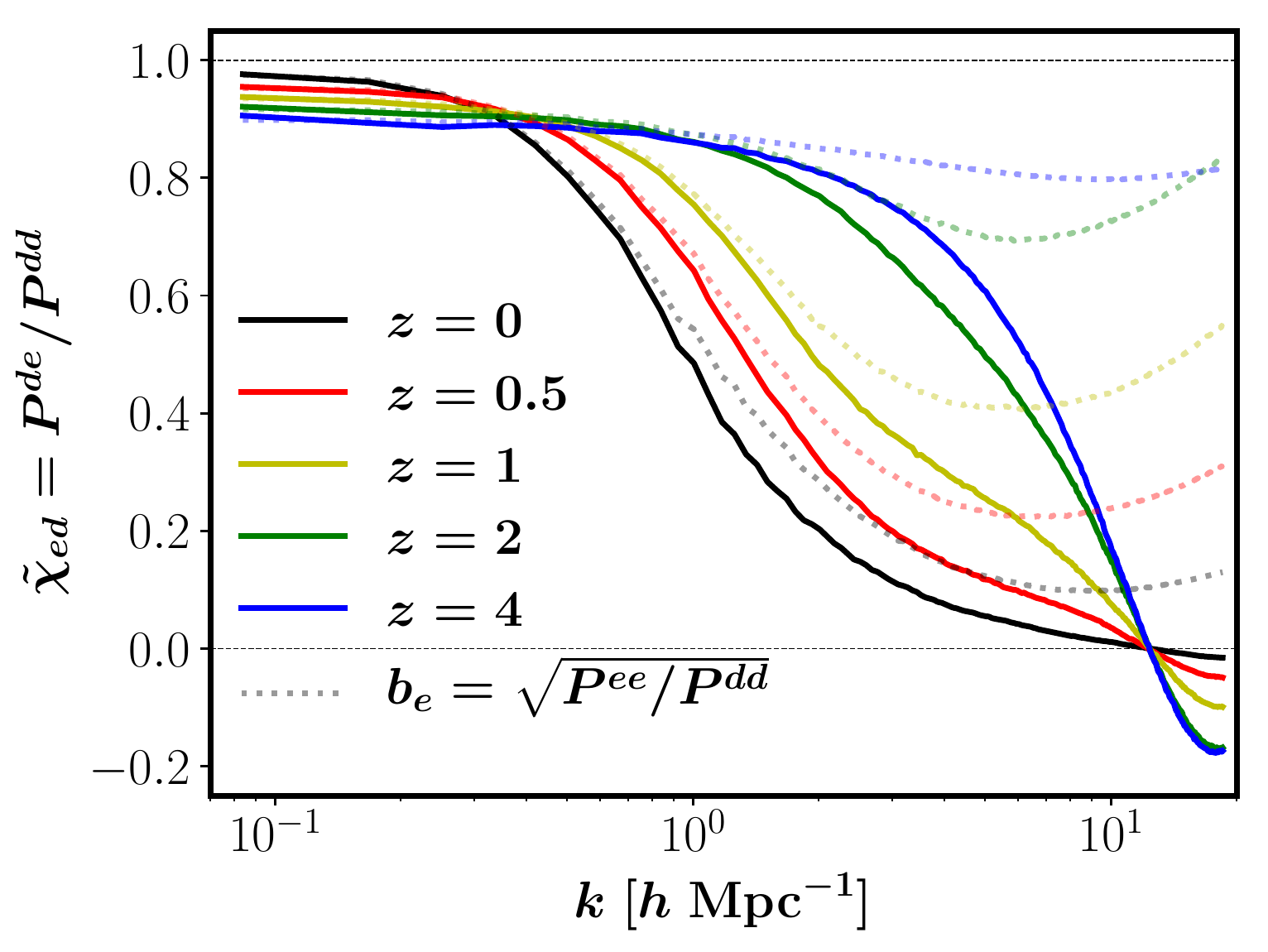}
    \includegraphics[scale=0.5,trim = 0 0 0 0, clip=true]{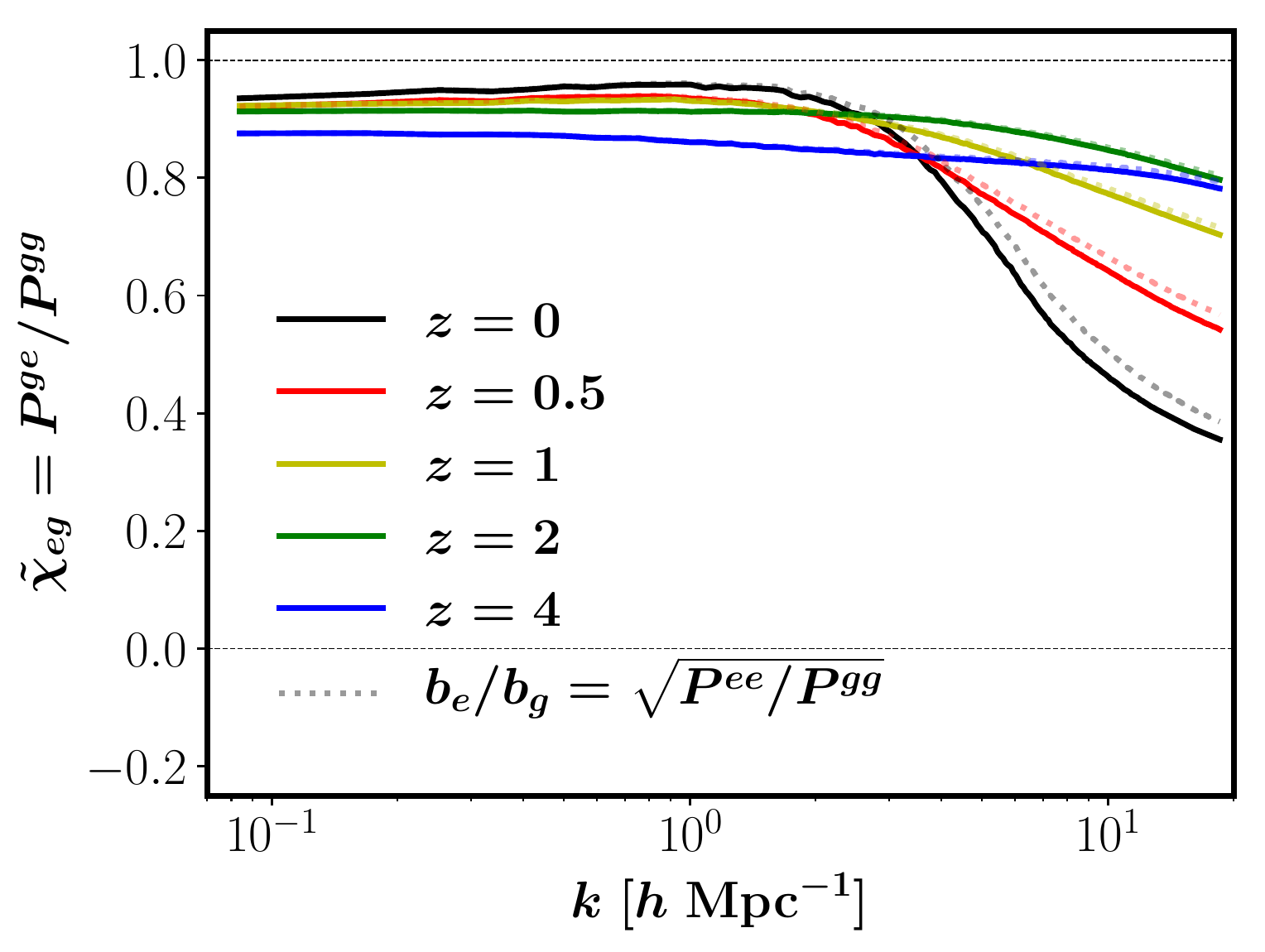}
     \includegraphics[scale=0.5,trim = 0 0 0 0, clip=true]{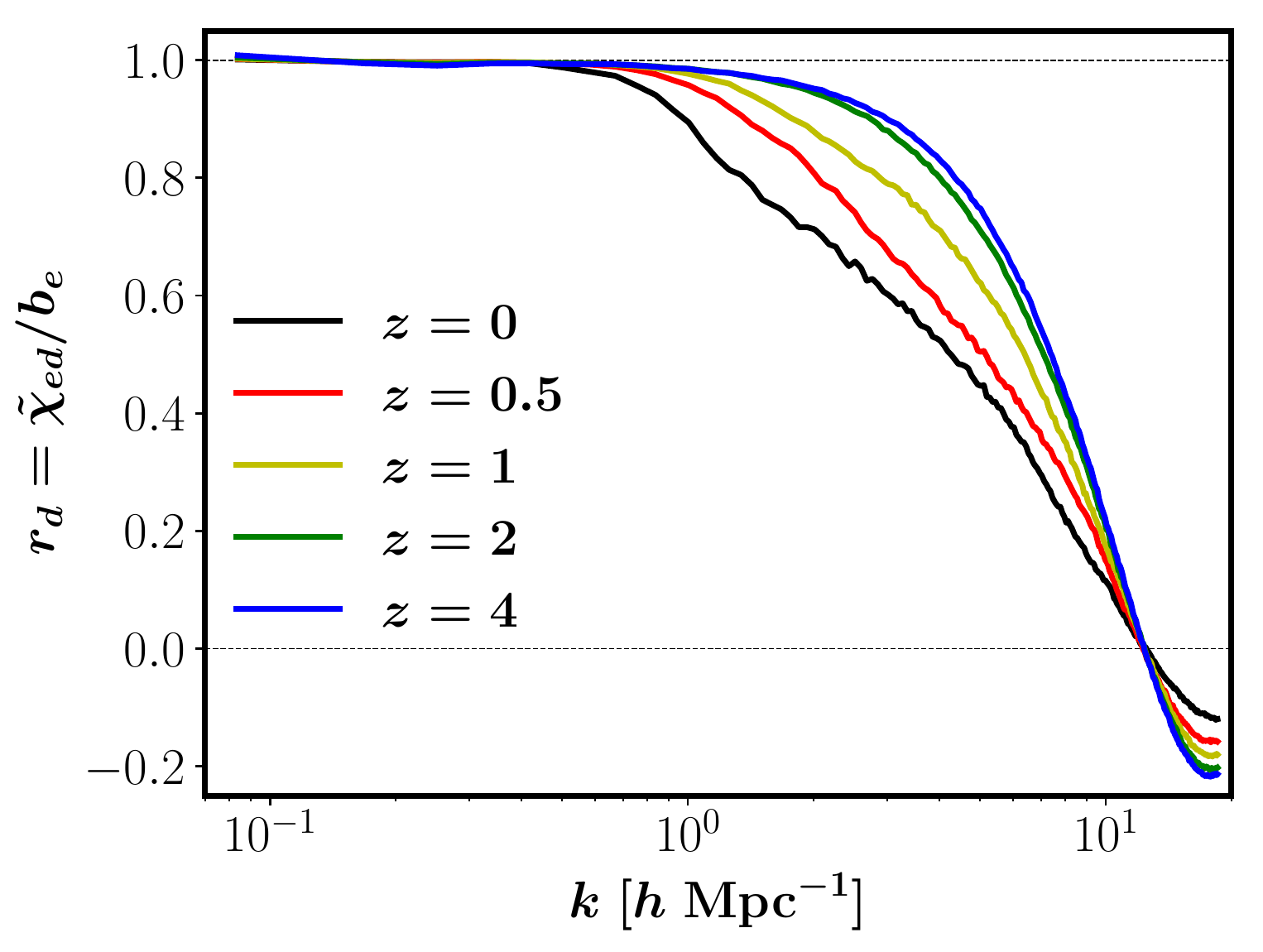}
     \caption{Fourier transforms of the susceptibilities $\tilde{\chi}_{gd}, \, \tilde{\chi}_{ed} $ and $ \tilde{\chi}_{eg}$ are compared  in panels 1, 2 and 3 with the biases $b_g, \, b_e $ and $b_e/b_g$ defined by ratios of the squares-roots of autocorrelation power spectra. These show the expected similarities at lower $k$, but also highlight the differences at high $k$. Panel 4 explicitly demonstrates this by plotting $r_d =  \tilde{\chi}_{de} \tilde{\chi}_{ed} =  \tilde{\chi}_{ed}^2 / b_e^2 $.  }
      \label{fig:chi}
      \end{center}
\end{figure*}

In the halo model expansion for the dark matter density, there are susceptibilities for each class of halo, giving the response of $\delta_d$ to $M_h n_h$, where $M_h$ is the halo mass and $n_h$ the halo number density for that class. In that case, the susceptibilities give the (suitably normalized) mean dark matter profile of the halo class. Usually scalings are done to take out  the dominant dependences on the overall size of the halo (or equivalently the overall mass). Thus these ideas are familiar. The Fourier transform $\tilde{\chi}_{eh} (k)$ is the electron density form factor of the halo.  

We can relate these ideas to what is done in practice to make mock maps of  the kSZ effect, e.g., in  the peak patch approach for creating deep all-sky halo catalogs \cite{1996LesHouches}, \cite{1996ApJS..103...63B}, and \cite{2017abs}, with analytic or measured susceptibilities used to "paint on"  the electron density profiles  around halos. The velocities for the halos are automatic outputs in the peak patch catalog, suitably smoothed over the cluster scale, which changes from cluster to cluster.  In \cite{2017abs} the susceptibilities are determined by measurement in the BBPS cosmic hydro simulations of \cite{2012ApJ...758...75B}, constructing  stacked profiles for halos with specified mass and redshift bins, each bin defining a halo-class.  The product of the electron density profile and the velocity allows for all-sky mocks of the  kSZ effect. The accuracy of the mock then depends upon the magnitude of the residual fluctuations about the mean-field linear-response expansion inherent in this ``halo model" for $n_e$. 

The Illustris simulations could also be used to calculate the electron density response for each of the halo bins, and be incorporated into the peak-patch simulator in the same way as the BBPS calculations have been. The resolution of Illustris was much superior to BBPS, which has allowed us to study here the influence of  galactic-scale baryonic physics that are not resolved in the BBPS simulations, although star and AGN feedback were incorporated. On the other hand, the Illustris box was smaller, and only one box was run, limiting statistical inferences. Thus in Illustris the rare-mass clusters are fewer, exacerbated because there was only one realization. (BBPS ran a (small) ensemble of simulations.) While we await more and larger Illustris and other high resolution simulations for susceptibility measurements, we shall content ourselves here with  $\chi_{ed}$ measurements, estimated by cross-correlations weighted by the inverse of the self-clustering correlation matrix for dark matter $<\delta_e \delta_d> <\delta_d\delta_d> ^{-1}$: 
\bea  
\tilde{\chi}_{ed} (k) = P^{ed}_{\delta\delta}(k)/P^{d}_{\delta\delta}(k).  
\eea 
For halos  $\tilde{\chi}_{dh} (k)$ approaches a mass-weighted anti-bias $b_{dh0}$ for the halo-class in question at low $k$, $\delta_d \sim \delta_h /b_{hd0}$, with  $b_{hd0} = \tilde{\chi}_{hd} (k\rightarrow 0)$ the usual mass-halo dependent bias of halos relative to dark matter density, with $b_{hd0}>1$. By analogy,  we can think of the dimensionless $\tilde{\chi}_{ed} (k)$ as the scale-dependent bias of electron density relative to dark matter density. 
 
 In the susceptibility language, $b^2_{e} = P^{e}_{\delta\delta}/P^{d}_{\delta\delta} =  \tilde{\chi}_{ed} (k) /\tilde{\chi}_{de} (k)$, hence asymptotically approaches $b_{ed0}/b_{de0}$.  Although naive biasing prescriptions would suggest this approaches $b_{ed0}^2$, in practice it is not quite true: we do not have  $ \tilde{\chi}_{de} (k) = 1/ \tilde{\chi}_{ed} (k) $, so $b_{e} $ also does not reduce to the biasing factor of $e$ relative to $d$. However as we  show in Fig.~\ref{fig:chi} the basic features of $b_{e}$ look rather similar to those of $\tilde\chi_{ed}$, though with different high $k$ suppression behavior. 
 
\subsection{Connected Term in Free Electron Transverse Momentum Power Spectrum }

$P_{q_\perp, uc}$ (Eq.~\ref{eq:Pq_unconnected}) was often assumed to represent the complete $P_{q_\perp}$ in  previous literature. SRN12 and \citet{2004MNRAS.347.1224Z} showed that evaluating Equation~(\ref{eq:Pq_unconnected}) using density power spectrum from simulations and directly computing $P_{q_\perp}$ gave nearly similar results. P16 recently revisited that comparison with a pure dark matter simulation and found that $P_{q_\perp}$ can be larger than $P_{q_\perp, uc}$ up to $\sim30\%$ at $k\gtrsim 1~h~\rm{Mpc}^{-1}$ due to the connected term in $P_{q_\perp}$ arising in the nonlinear regime. We shall check on this issue for the free electron momentum field here.

To measure $P_{q_\perp}$ directly from Illustris, we first assign the velocity-weighted electron number of each particle to a nearest grid in a $512^3$ mesh. Taking the power spectrum of this mesh would however underestimate $P_{q_\perp}$ because it misses velocity modes larger than the size of the simulation box ($75~h^{-1}~\rm{Mpc}$) and we thus need to correct for the missing velocity modes not included in the box. We describe how we estimate the correction for those missing modes in the Appendix and show the expression in Equation~(\ref{eq:missing_power}). In the rest of the paper, we shall use the missing-power-correction-added power spectrum in the analysis and denote it as $P_{q_\perp,tot}$.

\begin{figure}
  \begin{center}
    \includegraphics[scale=0.5]{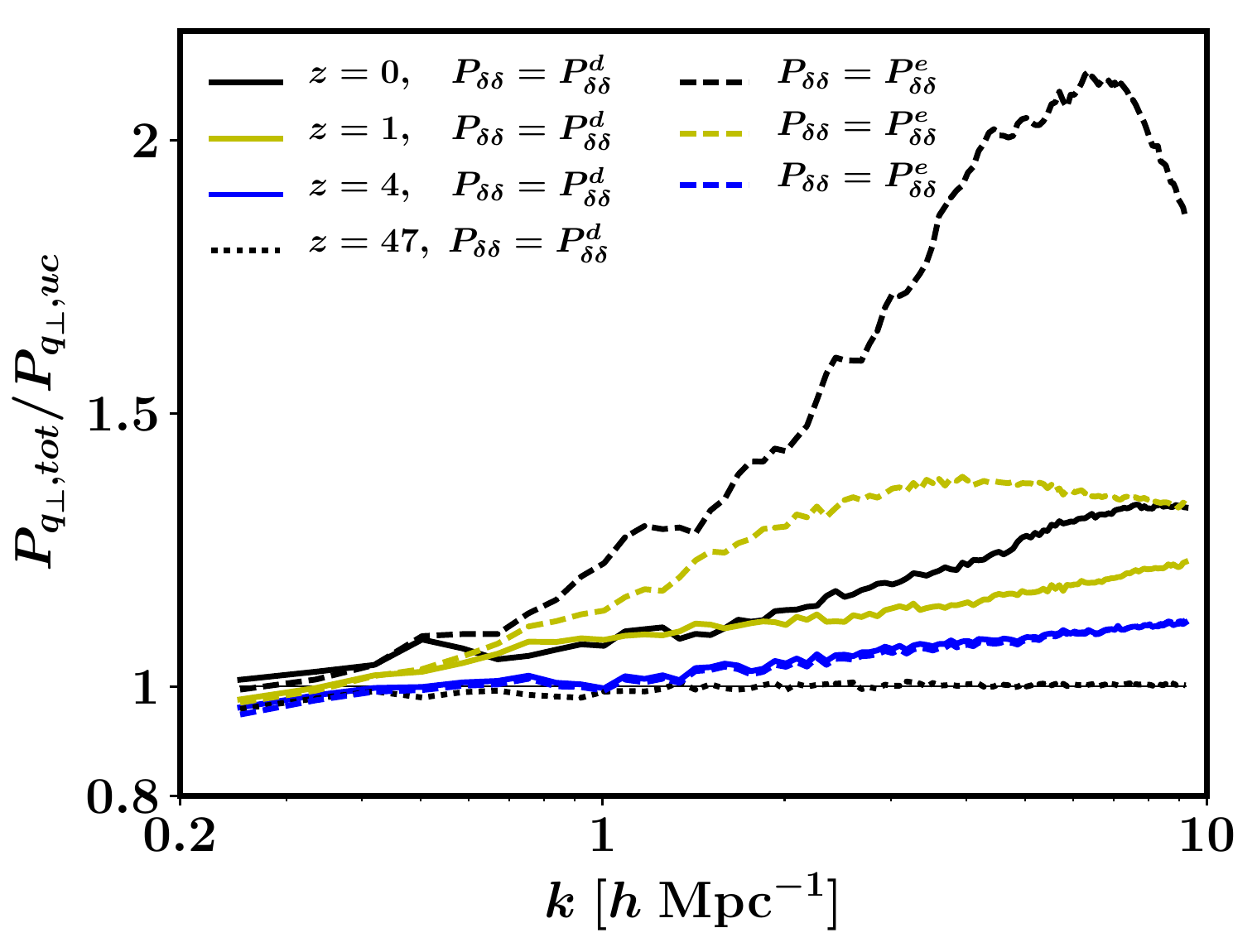}
  \caption{ Ratio between the missing-power-corrected transverse momentum power spectrum from the simulation ($P_{q_\perp,tot}$) and the unconnected term ($P_{q_\perp,uc}$). The solid lines describe the dark matter momentum field at $z = 0$ (black), 1 (yellow), and 4 (blue) and the dashed lines describe the free electron momentum field at the same redshifts. The black dotted line in addition describes the dark matter case at $z=47$.}
  \label{fig:Pqper_compare}
  \end{center}
\end{figure}

In Figure~\ref{fig:Pqper_compare}, we plot the ratio of $P_{q_\perp,tot}$, and the unconnected part of the spectrum, $P_{q_\perp,uc}$, to see if $P_{q_\perp,uc}$ can fully explains the spectrum. $P_{q_\perp,uc}$ is calculated by evaluating Equation~(\ref{eq:Pq_unconnected}) with the density and velocity power spectra of Illustris. We focus on the higher $k>0.5~h~\rm {Mpc}^{-1}$ wavenumbers where the kSZ signal is quite relevant and the number of plane-waves in the simulation box are sufficient for good statistically sampling.

At $z = 47$ where the dark matter density and velocity fluctuations are in the linear regime, $P_{q_\perp,tot}$ agrees with $P_{q_\perp,uc}$ within a percent. At $z\le4$, $P_{q_\perp,tot}$ shows some excess power at $k\gtrsim1~h~\rm{Mpc}^{-1}$. The excess grows monotonically toward low-$z$. At $z\le1$, the growth of the excess power is much faster for the free electron case than in the dark matter case, which we attribute to the baryonic physics adding more non-Gaussianity to the free electron density field. We interpret this excess power, 
\bea \label{eq:Pqc}
P_{q_\perp,c} \equiv P_{q_\perp,tot}-P_{q_\perp,uc}, 
\eea
as the contribution from the connected term. We note that this is only a conservative lower-bound since we have  made no missing power correction for $P_{q_\perp,c}$. If we allow for the same fraction of missing power as in $P_{q_\perp,uc}$, the estimated $P_{q_\perp,c}$ would nearly double, but the way the waves enter into the 4-point function precludes that being an accurate correction. We discuss how much $P_{q_\perp,c}$ adds to the integrated kSZ signal in the next section. 

\subsection{Angular Power Spectrum of the kSZ effect}

\begin{figure}[t]
  \begin{center}
    \vspace{0.2cm}
    \includegraphics[scale=0.55,trim = 0 0 0 26, clip=true]{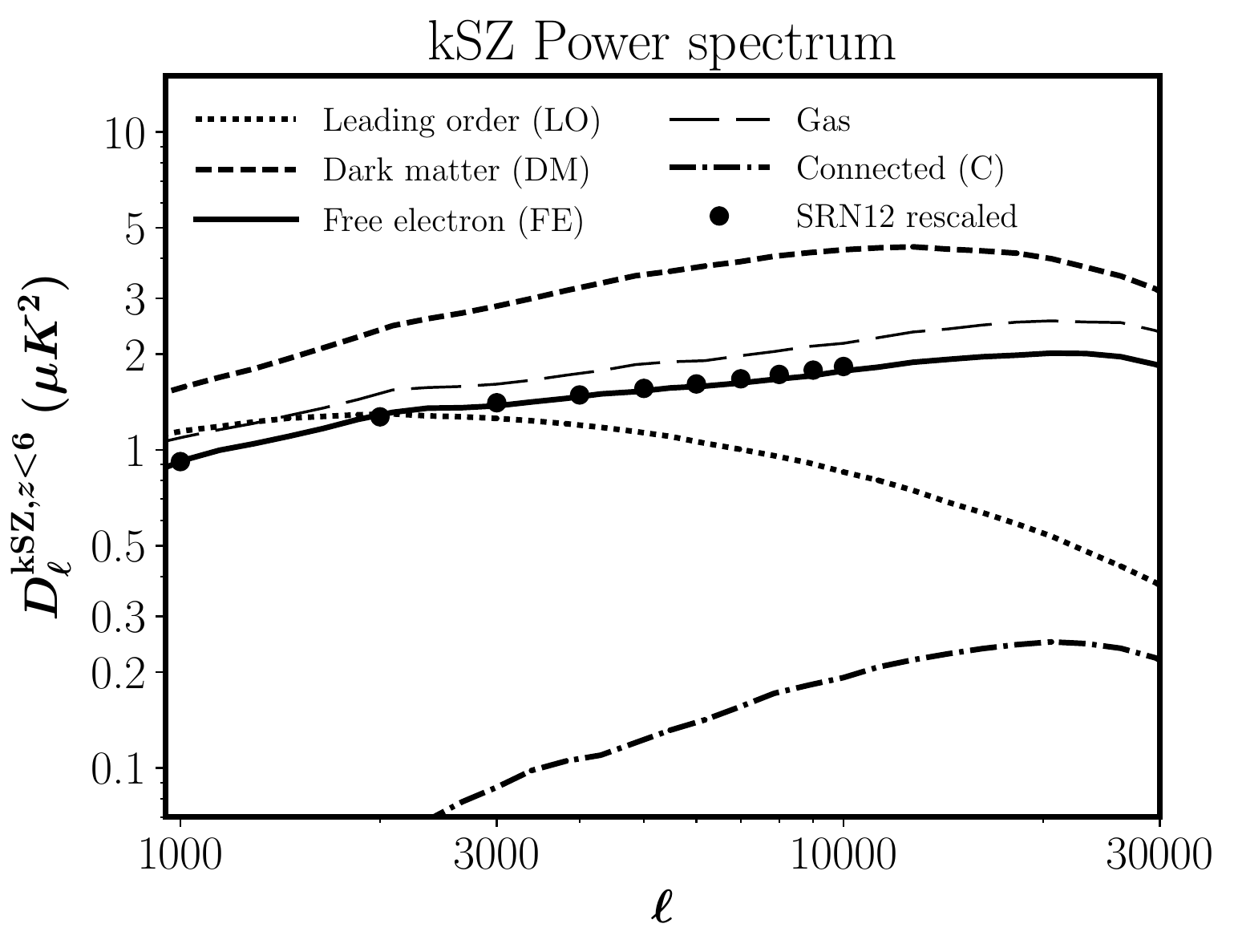} 
  \caption{Angular power spectrum of the kSZ effect from the post-reionization ($z<6$) epoch for the models considered in this work. The LO, DM, FE, Gas, and C models defined in the text are shown in dotted, short-dashed, solid, long-dashed, and dotted-dashed lines, respectively. Additionally, we show the prediction of the CSF model or SRN12 for our cosmological parameters. }
  \label{fig:kSZ}
  \end{center}
\end{figure}

We evaluate Equation~(\ref{eq:kSZ_Cl}) using $P_{q_\perp,tot}$ to calculate the kSZ signal from the post-reionization era, $\mathcal{D}^{{\rm kSZ},z<6}_{\ell}$, for several different models and present it in Figure~\ref{fig:kSZ}. We also list $\mathcal{D}^{{\rm kSZ},z<6}_{\ell}$ at $\ell =3000$ and $10000$ in Table~\ref{table: fitting}. For models that do not use the free electron density power spectrum, we assume an instantaneous helium reionization at $z=3$ (i.e. helium is singly ionized at $3\le z \le6$ and doubly ionized at $z<3$). We describe the models below.

For the leading order (LO) model, we set  $f_e=1$ and use $P_{q_\perp,uc}$ given by linear $P_{\delta\delta}$ and $P_{vv}$. This model excludes all the nonlinear effects in the momentum field and is often referred to as the Ostriker-Vishniac spectrum in the literature \citep{1987MNRAS.227P..33E,1987ApJ...322..597V} and gives $\mathcal{D}^{{\rm kSZ},z<6}_{\ell}=1.26 - 0.85~\mu K^2$ at $\ell=3000-10000$.

For the DM model, we set $f_e=1$ and use the dark matter momentum power spectrum from the simulation. The distribution of dark matter is not strongly  affected by the baryonic physics except at short scales. We can thus assume this model predominantly shows the effect of nonlinear growth of structure and not the effect of baryonic physics. Compared to in the LO model, the kSZ signal is enhanced to $3.83 - 4.26~\mu K^2$ at $\ell = 3000 - 10000$.

The Gas and FE models shown denote the cases for which we use gas and free electron momentum power spectrum, respectively. We use $f_g$ instead of $f_e$ for the Gas model and the true value of $f_e$ for the FE model. The FE model is the main result of this work. Compared to in the DM model, the kSZ signal at $\ell=3000-10000$ is suppressed to $\sim 60-50\%$  in the Gas model and $\sim 50-40\%$ in the FE model. Considering that, the all-wavenumber suppression factor $f^2b^2_0$ is $\sim 0.9$ and $0.7$ for the Gas and FE models (See Fig.~\ref{fig:fb}), whereas the high-$k$ suppression seems to give an extra $\sim30 (40)\%$ suppression for $\ell=3000\, (10000)$. The difference between the Gas and FE models is thus attributable to the difference between $f_g^2b_{g0}^2$ and $f_e^2 b_{e0}^2$.

For the FE model, we show how much is  contributed by the connected term by computing the kSZ signal from $P_{q_\perp, c}$. We refer to this case as the `C' model. The result shows that the connected term contributes $6-11\%$ to the signal at $\ell =3000-10000$.  As mentioned in the previous section, the connected term may have been underestimated due to the finite box-size of the simulation, possibly even by factor of a few.

In Table~\ref{table: fitting} we provide $\mathcal{D}^{{\rm kSZ},z<6}_{\ell=3000}$ from other studies rescaled to our cosmology. SRN12 used $z_{\rm rei}=10$ as the end of reionization redshift and gave $\mathcal{D}^{{\rm kSZ},z<10}_{\ell=3000}=2.19~\mu K^2$ and $\mathcal{D}^{{\rm kSZ},z<10}_{\ell=10000}=2.52~\mu K^2$. Scaling their result for $z_{\rm rei}=6$ with the power-law indices they provided\footnote{See Table 3 of their work.}  gives $\mathcal{D}^{{\rm kSZ},z<6}_{\ell=3000}=1.40~\mu K^2$ and $\mathcal{D}^{s{\rm kSZ},z<6}_{\ell=10000}=1.83~\mu K^2$. This is in a very good agreement with our FE model (also see their power spectrum in Fig.~\ref{fig:kSZ}). We also show similarly rescaled values from the standard model of \citet[][TBO11]{2011ApJ...727...94T}, the AGN model of \citet[][B10]{2010ApJ...725...91B}, \citet[][DKS16]{2016MNRAS.463.1797D}, and \citet[][RVB17]{2017MNRAS.tmp..176R}. The results give $\mathcal{D}^{{\rm kSZ},z<6}_{\ell=3000}=1.65,~1.75$, $2.1$, and $2.7~\mu K^2$, respectively. 

The $\ell$-dependence of $\mathcal{D}^{{\rm kSZ},z<6}_{\ell}$ is close to being power-law between $\ell = 3000$ and $10000$. Thus, we use
\bea \label{eq:fitting}
\mathcal{D}^{{\rm kSZ},z<6}_{\ell,\rm fit} =  \left(\frac{\ell}{3000}\right)^{\gamma_{\rm kSZ}} \mathcal{D}^{{\rm kSZ},z<6}_{\ell=3000}
\eea
with 
\bea
\gamma_{\rm kSZ} = \left. \log \left(\frac{\mathcal{D}^{{\rm kSZ},z<6}_{\ell=10000}}{\mathcal{D}^{{\rm kSZ},z<6}_{\ell=3000}}\right) \middle/ \log \left(\frac{10000}{3000}\right) \right.
\eea
for the fitting function. We list $\mathcal{D}^{{\rm kSZ},z<6}_{\ell=3000}$, $\mathcal{D}^{{\rm kSZ},z<6}_{\ell=10000}$, and $\gamma_{\rm kSZ}$ for each model in Table~\ref{table: fitting}. 

\begin{figure}
  \begin{center} 
  \vspace{0.2cm}
    \includegraphics[scale=0.55,trim = 0 0 0 47, clip=true]{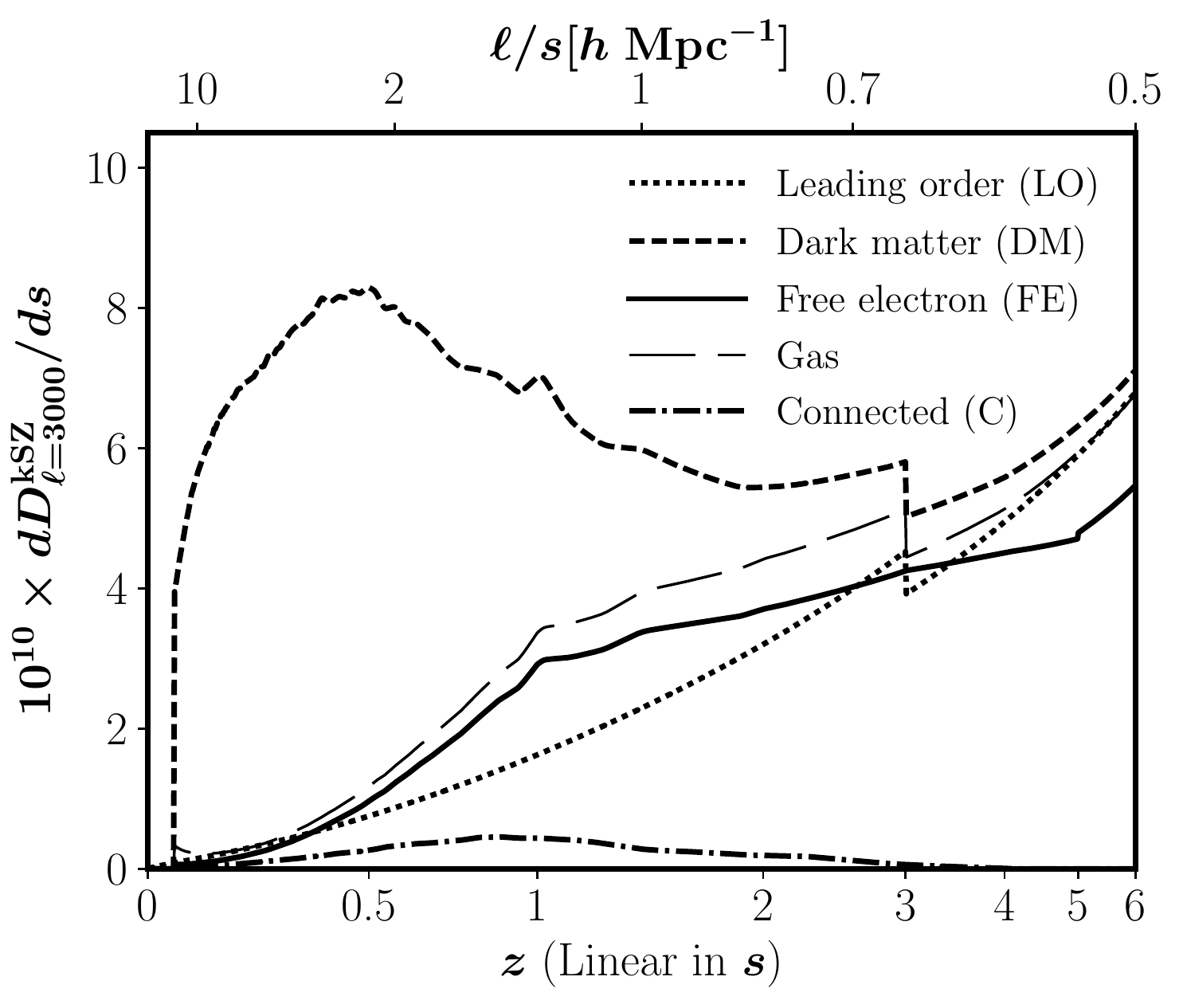} 
  \caption{Contribution to the kSZ signal from a given redshift, $d\mathcal{D}^{{\rm kSZ}}_{\ell=3000}/d\log{z}$, for several models. Each line type corresponds to the same kSZ model as in Figure~\ref{fig:kSZ}.  The $x$-axis is linear in the comoving distance, $s$, although it is labeled in redshift in the lower side and the contributing wavenumber, $k=\ell/s$ in the upper side. The area under each curve is proportional to the integrated kSZ signal of each model.}
  \label{fig:kSZ_int}
  \end{center}
\end{figure}

In Figure~\ref{fig:kSZ_int}, we describe how the kSZ signal accumulates over redshift for each model by plotting $d\mathcal{D}^{{\rm kSZ}}_{\ell=3000}/d\log{z}$. The DM model in comparison to the LO model shows a huge enhancement at $z\lesssim2$ due to nonlinear growth of structure. This enhancement, however, is mostly canceled by baryonic suppression in the Gas and FE models, making the result similar to what the LO model gives. 
This makes the low-$z$ contribution relatively unimportant: only $\sim20\%$ of the signal comes from $z\le1$. Although the connected term grows large relative to the unconnected ones at $z\lesssim1$, the low-$z$ suppression keeps it to an $\lesssim 10\%$ level.

\section{Summary and Discussion}

The transverse momentum power spectrum $P_{q_\perp}$ that sources the kSZ signal in the CMB anisotropy has often been computed by substituting the free electron density power spectrum $P^e_{\delta\delta}$ acquired from simulation into Equation~(\ref{eq:Pq_unconnected}). According to the equation, $P_{q_\perp}$ at a given wavenumber depends sensitively on $P_{\delta\delta}$ at similar wavenumbers. Free electron density fluctuations at $z<6$ are heavily affected by nonlinear growth of structure and baryonic physics and modeling these nonlinear effects well are the key to estimating accurately the kSZ signal in the post-reionization regime. While nonlinear growth is known to be well described by the HALOFIT model, the baryonic effects are less understood due to the difficulty in implementing the physics into simulation, and are not simply amenable to semi-analytic treatments.

Although not perfectly accurate, Illustris provides a sample distribution of free electron at a high enough resolution in a large enough volume with relevant baryonic physics. Even after the end of reionization, a certain fraction of electron is locked in stars, black-holes, neutral gas, and helium. Since these are located preferentially in over-dense regions, the free electron density fluctuations are suppressed by the anti-bias as well as by the neutral fraction.  Another important effect is the feedback from AGNs blowing gas out of groups and clusters, resulting in a strong extra suppression on small scales. 

Since the dark matter density field is minimally affected by the baryonic effects, the window function $W^2_e=f^2_e[P^e_{\delta\delta}/P^d_{\delta\delta}]$ well separates the baryonic suppression effects on the kSZ signal. According to the window function from the Illustris simulation, the loss of gas and anti-bias suppresses the signal by $\sim 30\%$ at all scales. Then the extra small-scale suppression by AGN feedback further reduces the signal at $k\gtrsim1~h~\rm{Mpc}^{-1}$ at $z\lesssim1$. This small-scale suppression is more effective for higher $\ell$. The net suppression in the integrated post-reionization kSZ signal $\mathcal{D}^{{\rm kSZ},z<6}_\ell$ is at $50-60\%$ level at $\ell=3000-10000$.  We note that this small-scale suppression effect by AGN is likely over-estimated in Illustris. According  \cite{2014MNRAS.445..175G}, Illustris significantly undershoots the gas fraction in halos with $M\gtrsim 10^{13}M_\odot$ at $z\lesssim 1$ due to exaggeration in the AGN feedback prescription and these halos are an important contributor to the density power spectrum at $k\gtrsim1~h~\rm{Mpc}^{-1}$.

\begin{table}[t]
\caption{Power-law fitting (Eq.~\ref{eq:fitting}) parameters for $\mathcal{D}^{{\rm kSZ},z<6}_\ell$}
\begin{center} 
\begin{tabular}{@{}llllllllllllllllllll||}  
\hline
label & $\mathcal{D}^{{\rm kSZ},z<6}_{\ell=3000}$ $(\mu K^2)$  &  $\mathcal{D}^{{\rm kSZ},z<6}_{\ell=10000}$ $(\mu K^2)$ & $\gamma_{\rm kSZ}$  \\
\hline
LO          & 1.26 & 0.85 & -0.32 \\
DM          & 2.83 & 4.26 & 0.34 \\
Gas         & 1.61 & 2.17 & 0.25 \\
C             & 0.087& 0.192 & 0.66 \\
\bf{FE}\footnote{The main result for Illustris}           & \bf{1.38} & \bf{1.77} & \bf{0.21} \\
SRN12 (rescaled)   & 1.40 & 1.83 & 0.22 \\
TBO11 (rescaled)   & 1.65 & - &-\\
B10 (rescaled)   & 1.75 & - &-\\
DKS16 (rescaled)   & 2.1 & - &-\\
RVB17 (rescaled)   & 2.7 & 3.65 &0.25\\
\hline
\end{tabular}  
\end{center}
\label{table: fitting}
\end{table}

We also find the connected term, which is not accounted in Equation~(\ref{eq:Pq_unconnected}), rises steeply in the nonlinear regime when we directly compute $P_{q_\perp}$ from the simulation (not using Eq.~\ref{eq:Pq_unconnected}), corrected for the box-size effect, when we compare with the result of evaluating Equation~(\ref{eq:Pq_unconnected}). The importance of the connected term  in $P_{q_\perp}$  was already reported in P16 and \citet{2016ApJ...824..118A} for the pure dark matter momentum field and for the patchy ionized density field during EoR, respectively. Here, we not only confirm its significance for the free electron momentum field, but find that the connected term in the free electron case accounts for a much higher fraction in $P_{q_\perp}$ than in the pure dark matter case due to enhanced nonlinearity from the baryonic physics. At $k\gtrsim1~h~\rm{Mpc}^{-1}$ at $z\lesssim1$, the connected term constitutes $30\%$ or more of the $P_{q_\perp}$ and this fraction is only a conservative lower limit given that the connected term is not box-size corrected. This is a serious warning against relying on the unconnected term (Eq.~\ref{eq:Pq_unconnected}) for computing the kSZ signal. Due to the strong baryonic suppression at low-$z$, the connected term contribution to the integrated signal is only $6\%\, (12\%)$ at $\ell = 3000, (10000)$, but it may be a factor of few larger in larger simulations that are free from the box-size effect.

The resulting kSZ power spectrum is well described by a power-law fit, $\mathcal{D}^{{\rm kSZ},z<6}_\ell=1.38[\ell/3000]^{0.21}~\mu K^2$, at $\ell=3000-10000$. This is impressively close to the result of SRN12 that studied $W_e$ in the same context with a different simulation. However, the results from other similar recent kSZ studies are generally higher, varying from $1.7$ to $2.7~\mu K^2$ at $\ell=3000$ as listed in Table 1. 

Both with this Illustris simulation and the simulation used in SRN12, we expect the kSZ signal  to be underestimated. Our work with Illustris is subject to the exaggerated AGN feedback as mentioned above and SRN12 used the L60CSF simulation that is known to over-produce stars. The results of DKS16 ($2.1~\mu K^2$) and RVB17 ($2.7~\mu K^2$) are however too high to be explained simply by removing the high-$k$ suppression of the AGN feedback: applying the  $f^2_e b^2_{e0}\approx 0.7$ factor on the DM model and ignoring the high-$k$ suppression would yield $\sim2.0~\mu K^2$, which is still lower than those results. Another possibility is that the connected term is much larger in those works. The box-sizes used in RVB17 were 120, 240, and 480~$h^{-1}~\rm{Mpc}$ and the one used in DKS16 was nearly $900~h^{-1}~\rm{Mpc}$. For those larger box-sizes, the box-size effect on the connected term will be smaller. Also, a weaker AGN feedback may have resulted in the connected term  contributing more at low-$z$ than in Illustris. 

To clarify these discrepancies in the literature, we need to directly compare $W_e$, $f_e$, $b_e$, and the connected term in different Illustris-like high resolution gas simulations with varied feedback scenarios to calibrate the gastrophysical uncertainties in the kSZ predictions. The next round of high resolution large-sky fraction CMB experiments hold great promise for the observations, making such a concerted theoretical and simulational effort imperative. And if we can get to the point of making an accurate subtraction of the post-reionization signal from observations, we can hope to learn about the more gastrophysically complex epoch-of-reionization kSZ signal, and help to constrain EoR physics.

\section{Acknowledgments}

The authors thank E. Komatsu, S. Foreman, and M. McQuinn for helpful comments on this work. HP thanks the Illustris team for their assistance with using the Illustris data and C. Sabiu for his help with writing up this manuscript. The Illustris data used in this work was stored and analyzed on the GPC supercomputer at the SciNet HPC Consortium. SciNet is funded by: the Canada Foundation for Innovation under the auspices of Compute Canada; the Government of Ontario; Ontario Research Fund - Research Excellence; and the University of Toronto.

\bibliographystyle{apj}
\end{CJK}
\bibliography{reference}

\begin{thebibliography}{57}
\expandafter\ifx\csname natexlab\endcsname\relax\def\natexlab#1{#1}\fi
\expandafter\ifx\csname bibnamefont\endcsname\relax
  \def\bibnamefont#1{#1}\fi
\expandafter\ifx\csname bibfnamefont\endcsname\relax
  \def\bibfnamefont#1{#1}\fi
\expandafter\ifx\csname citenamefont\endcsname\relax
  \def\citenamefont#1{#1}\fi
\expandafter\ifx\csname url\endcsname\relax
  \def\url#1{\texttt{#1}}\fi
\expandafter\ifx\csname urlprefix\endcsname\relax\def\urlprefix{URL }\fi
\providecommand{\bibinfo}[2]{#2}
\providecommand{\eprint}[2][]{\url{#2}}

\bibitem[{\citenamefont{{Sunyaev} and {Zeldovich}}(1972)}]{1972CoASP...4..173S}
\bibinfo{author}{\bibfnamefont{R.~A.} \bibnamefont{{Sunyaev}}}
  \bibnamefont{and} \bibinfo{author}{\bibfnamefont{Y.~B.}
  \bibnamefont{{Zeldovich}}}, \bibinfo{journal}{Comments on Astrophysics and
  Space Physics} \textbf{\bibinfo{volume}{4}}, \bibinfo{pages}{173}
  (\bibinfo{year}{1972}).

\bibitem[{\citenamefont{{Sunyaev} and {Zeldovich}}(1980)}]{1980MNRAS.190..413S}
\bibinfo{author}{\bibfnamefont{R.~A.} \bibnamefont{{Sunyaev}}}
  \bibnamefont{and} \bibinfo{author}{\bibfnamefont{I.~B.}
  \bibnamefont{{Zeldovich}}}, \bibinfo{journal}{\mnras}
  \textbf{\bibinfo{volume}{190}}, \bibinfo{pages}{413} (\bibinfo{year}{1980}).

\bibitem[{\citenamefont{{Komatsu} and {Seljak}}(2002)}]{2002MNRAS.336.1256K}
\bibinfo{author}{\bibfnamefont{E.}~\bibnamefont{{Komatsu}}} \bibnamefont{and}
  \bibinfo{author}{\bibfnamefont{U.}~\bibnamefont{{Seljak}}},
  \bibinfo{journal}{\mnras} \textbf{\bibinfo{volume}{336}},
  \bibinfo{pages}{1256} (\bibinfo{year}{2002}), \eprint{astro-ph/0205468}.

\bibitem[{\citenamefont{{Abazajian} et~al.}(2016)\citenamefont{{Abazajian},
  {Adshead}, {Ahmed}, {Allen}, {Alonso}, {Arnold}, {Baccigalupi}, {Bartlett},
  {Battaglia}, {Benson} et~al.}}]{2016arXiv161002743A}
\bibinfo{author}{\bibfnamefont{K.~N.} \bibnamefont{{Abazajian}}},
  \bibinfo{author}{\bibfnamefont{P.}~\bibnamefont{{Adshead}}},
  \bibinfo{author}{\bibfnamefont{Z.}~\bibnamefont{{Ahmed}}},
  \bibinfo{author}{\bibfnamefont{S.~W.} \bibnamefont{{Allen}}},
  \bibinfo{author}{\bibfnamefont{D.}~\bibnamefont{{Alonso}}},
  \bibinfo{author}{\bibfnamefont{K.~S.} \bibnamefont{{Arnold}}},
  \bibinfo{author}{\bibfnamefont{C.}~\bibnamefont{{Baccigalupi}}},
  \bibinfo{author}{\bibfnamefont{J.~G.} \bibnamefont{{Bartlett}}},
  \bibinfo{author}{\bibfnamefont{N.}~\bibnamefont{{Battaglia}}},
  \bibinfo{author}{\bibfnamefont{B.~A.} \bibnamefont{{Benson}}},
  \bibnamefont{et~al.}, \bibinfo{journal}{ArXiv e-prints}
  (\bibinfo{year}{2016}), \eprint{1610.02743}.

\bibitem[{\citenamefont{{Hern{\'a}ndez-Monteagudo}
  et~al.}(2015)\citenamefont{{Hern{\'a}ndez-Monteagudo}, {Ma}, {Kitaura},
  {Wang}, {G{\'e}nova-Santos}, {Mac{\'{\i}}as-P{\'e}rez}, and
  {Herranz}}}]{2015PhRvL.115s1301H}
\bibinfo{author}{\bibfnamefont{C.}~\bibnamefont{{Hern{\'a}ndez-Monteagudo}}},
  \bibinfo{author}{\bibfnamefont{Y.-Z.} \bibnamefont{{Ma}}},
  \bibinfo{author}{\bibfnamefont{F.~S.} \bibnamefont{{Kitaura}}},
  \bibinfo{author}{\bibfnamefont{W.}~\bibnamefont{{Wang}}},
  \bibinfo{author}{\bibfnamefont{R.}~\bibnamefont{{G{\'e}nova-Santos}}},
  \bibinfo{author}{\bibfnamefont{J.}~\bibnamefont{{Mac{\'{\i}}as-P{\'e}rez}}},
  \bibnamefont{and}
  \bibinfo{author}{\bibfnamefont{D.}~\bibnamefont{{Herranz}}},
  \bibinfo{journal}{Physical Review Letters} \textbf{\bibinfo{volume}{115}},
  \bibinfo{eid}{191301} (\bibinfo{year}{2015}), \eprint{1504.04011}.

\bibitem[{\citenamefont{{Schaan} et~al.}(2016)\citenamefont{{Schaan},
  {Ferraro}, {Vargas-Maga{\~n}a}, {Smith}, {Ho}, {Aiola}, {Battaglia}, {Bond},
  {De Bernardis}, {Calabrese} et~al.}}]{2016PhRvD..93h2002S}
\bibinfo{author}{\bibfnamefont{E.}~\bibnamefont{{Schaan}}},
  \bibinfo{author}{\bibfnamefont{S.}~\bibnamefont{{Ferraro}}},
  \bibinfo{author}{\bibfnamefont{M.}~\bibnamefont{{Vargas-Maga{\~n}a}}},
  \bibinfo{author}{\bibfnamefont{K.~M.} \bibnamefont{{Smith}}},
  \bibinfo{author}{\bibfnamefont{S.}~\bibnamefont{{Ho}}},
  \bibinfo{author}{\bibfnamefont{S.}~\bibnamefont{{Aiola}}},
  \bibinfo{author}{\bibfnamefont{N.}~\bibnamefont{{Battaglia}}},
  \bibinfo{author}{\bibfnamefont{J.~R.} \bibnamefont{{Bond}}},
  \bibinfo{author}{\bibfnamefont{F.}~\bibnamefont{{De Bernardis}}},
  \bibinfo{author}{\bibfnamefont{E.}~\bibnamefont{{Calabrese}}},
  \bibnamefont{et~al.}, \bibinfo{journal}{\prd} \textbf{\bibinfo{volume}{93}},
  \bibinfo{eid}{082002} (\bibinfo{year}{2016}), \eprint{1510.06442}.

\bibitem[{\citenamefont{{Hill} et~al.}(2016)\citenamefont{{Hill}, {Ferraro},
  {Battaglia}, {Liu}, and {Spergel}}}]{2016PhRvL.117e1301H}
\bibinfo{author}{\bibfnamefont{J.~C.} \bibnamefont{{Hill}}},
  \bibinfo{author}{\bibfnamefont{S.}~\bibnamefont{{Ferraro}}},
  \bibinfo{author}{\bibfnamefont{N.}~\bibnamefont{{Battaglia}}},
  \bibinfo{author}{\bibfnamefont{J.}~\bibnamefont{{Liu}}}, \bibnamefont{and}
  \bibinfo{author}{\bibfnamefont{D.~N.} \bibnamefont{{Spergel}}},
  \bibinfo{journal}{Physical Review Letters} \textbf{\bibinfo{volume}{117}},
  \bibinfo{eid}{051301} (\bibinfo{year}{2016}), \eprint{1603.01608}.

\bibitem[{\citenamefont{{Soergel} et~al.}(2016)\citenamefont{{Soergel},
  {Flender}, {Story}, {Bleem}, {Giannantonio}, {Efstathiou}, {Rykoff},
  {Benson}, {Crawford}, {Dodelson} et~al.}}]{2016MNRAS.461.3172S}
\bibinfo{author}{\bibfnamefont{B.}~\bibnamefont{{Soergel}}},
  \bibinfo{author}{\bibfnamefont{S.}~\bibnamefont{{Flender}}},
  \bibinfo{author}{\bibfnamefont{K.~T.} \bibnamefont{{Story}}},
  \bibinfo{author}{\bibfnamefont{L.}~\bibnamefont{{Bleem}}},
  \bibinfo{author}{\bibfnamefont{T.}~\bibnamefont{{Giannantonio}}},
  \bibinfo{author}{\bibfnamefont{G.}~\bibnamefont{{Efstathiou}}},
  \bibinfo{author}{\bibfnamefont{E.}~\bibnamefont{{Rykoff}}},
  \bibinfo{author}{\bibfnamefont{B.~A.} \bibnamefont{{Benson}}},
  \bibinfo{author}{\bibfnamefont{T.}~\bibnamefont{{Crawford}}},
  \bibinfo{author}{\bibfnamefont{S.}~\bibnamefont{{Dodelson}}},
  \bibnamefont{et~al.}, \bibinfo{journal}{\mnras}
  \textbf{\bibinfo{volume}{461}}, \bibinfo{pages}{3172} (\bibinfo{year}{2016}),
  \eprint{1603.03904}.

\bibitem[{\citenamefont{{De Bernardis} et~al.}(2017)\citenamefont{{De
  Bernardis}, {Aiola}, {Vavagiakis}, {Battaglia}, {Niemack}, {Beall}, {Becker},
  {Bond}, {Calabrese}, {Cho} et~al.}}]{2017JCAP...03..008D}
\bibinfo{author}{\bibfnamefont{F.}~\bibnamefont{{De Bernardis}}},
  \bibinfo{author}{\bibfnamefont{S.}~\bibnamefont{{Aiola}}},
  \bibinfo{author}{\bibfnamefont{E.~M.} \bibnamefont{{Vavagiakis}}},
  \bibinfo{author}{\bibfnamefont{N.}~\bibnamefont{{Battaglia}}},
  \bibinfo{author}{\bibfnamefont{M.~D.} \bibnamefont{{Niemack}}},
  \bibinfo{author}{\bibfnamefont{J.}~\bibnamefont{{Beall}}},
  \bibinfo{author}{\bibfnamefont{D.~T.} \bibnamefont{{Becker}}},
  \bibinfo{author}{\bibfnamefont{J.~R.} \bibnamefont{{Bond}}},
  \bibinfo{author}{\bibfnamefont{E.}~\bibnamefont{{Calabrese}}},
  \bibinfo{author}{\bibfnamefont{H.}~\bibnamefont{{Cho}}},
  \bibnamefont{et~al.}, \bibinfo{journal}{JCAP} \textbf{\bibinfo{volume}{3}},
  \bibinfo{eid}{008} (\bibinfo{year}{2017}), \eprint{1607.02139}.

\bibitem[{\citenamefont{{Battaglia} et~al.}(2017)\citenamefont{{Battaglia},
  {Ferraro}, {Schaan}, and {Spergel}}}]{2017arXiv170505881B}
\bibinfo{author}{\bibfnamefont{N.}~\bibnamefont{{Battaglia}}},
  \bibinfo{author}{\bibfnamefont{S.}~\bibnamefont{{Ferraro}}},
  \bibinfo{author}{\bibfnamefont{E.}~\bibnamefont{{Schaan}}}, \bibnamefont{and}
  \bibinfo{author}{\bibfnamefont{D.}~\bibnamefont{{Spergel}}},
  \bibinfo{journal}{ArXiv e-prints}  (\bibinfo{year}{2017}),
  \eprint{1705.05881}.

\bibitem[{\citenamefont{{Mueller}
  et~al.}(2015{\natexlab{a}})\citenamefont{{Mueller}, {de Bernardis}, {Bean},
  and {Niemack}}}]{2015ApJ...808...47M}
\bibinfo{author}{\bibfnamefont{E.-M.} \bibnamefont{{Mueller}}},
  \bibinfo{author}{\bibfnamefont{F.}~\bibnamefont{{de Bernardis}}},
  \bibinfo{author}{\bibfnamefont{R.}~\bibnamefont{{Bean}}}, \bibnamefont{and}
  \bibinfo{author}{\bibfnamefont{M.~D.} \bibnamefont{{Niemack}}},
  \bibinfo{journal}{\apj} \textbf{\bibinfo{volume}{808}}, \bibinfo{eid}{47}
  (\bibinfo{year}{2015}{\natexlab{a}}), \eprint{1408.6248}.

\bibitem[{\citenamefont{{Sugiyama} et~al.}(2017)\citenamefont{{Sugiyama},
  {Okumura}, and {Spergel}}}]{2017JCAP...01..057S}
\bibinfo{author}{\bibfnamefont{N.~S.} \bibnamefont{{Sugiyama}}},
  \bibinfo{author}{\bibfnamefont{T.}~\bibnamefont{{Okumura}}},
  \bibnamefont{and} \bibinfo{author}{\bibfnamefont{D.~N.}
  \bibnamefont{{Spergel}}}, \bibinfo{journal}{JCAP}
  \textbf{\bibinfo{volume}{1}}, \bibinfo{eid}{057} (\bibinfo{year}{2017}),
  \eprint{1606.06367}.

\bibitem[{\citenamefont{{Mueller}
  et~al.}(2015{\natexlab{b}})\citenamefont{{Mueller}, {de Bernardis}, {Bean},
  and {Niemack}}}]{2015PhRvD..92f3501M}
\bibinfo{author}{\bibfnamefont{E.-M.} \bibnamefont{{Mueller}}},
  \bibinfo{author}{\bibfnamefont{F.}~\bibnamefont{{de Bernardis}}},
  \bibinfo{author}{\bibfnamefont{R.}~\bibnamefont{{Bean}}}, \bibnamefont{and}
  \bibinfo{author}{\bibfnamefont{M.~D.} \bibnamefont{{Niemack}}},
  \bibinfo{journal}{\prd} \textbf{\bibinfo{volume}{92}}, \bibinfo{eid}{063501}
  (\bibinfo{year}{2015}{\natexlab{b}}), \eprint{1412.0592}.

\bibitem[{\citenamefont{{Alonso} et~al.}(2016)\citenamefont{{Alonso}, {Louis},
  {Bull}, and {Ferreira}}}]{2016PhRvD..94d3522A}
\bibinfo{author}{\bibfnamefont{D.}~\bibnamefont{{Alonso}}},
  \bibinfo{author}{\bibfnamefont{T.}~\bibnamefont{{Louis}}},
  \bibinfo{author}{\bibfnamefont{P.}~\bibnamefont{{Bull}}}, \bibnamefont{and}
  \bibinfo{author}{\bibfnamefont{P.~G.} \bibnamefont{{Ferreira}}},
  \bibinfo{journal}{\prd} \textbf{\bibinfo{volume}{94}}, \bibinfo{eid}{043522}
  (\bibinfo{year}{2016}), \eprint{1604.01382}.

\bibitem[{\citenamefont{{Cooray}}(2004)}]{2004PhRvD..70f3509C}
\bibinfo{author}{\bibfnamefont{A.}~\bibnamefont{{Cooray}}},
  \bibinfo{journal}{\prd} \textbf{\bibinfo{volume}{70}}, \bibinfo{eid}{063509}
  (\bibinfo{year}{2004}), \eprint{astro-ph/0405528}.

\bibitem[{\citenamefont{{Salvaterra} et~al.}(2005)\citenamefont{{Salvaterra},
  {Ciardi}, {Ferrara}, and {Baccigalupi}}}]{2005MNRAS.360.1063S}
\bibinfo{author}{\bibfnamefont{R.}~\bibnamefont{{Salvaterra}}},
  \bibinfo{author}{\bibfnamefont{B.}~\bibnamefont{{Ciardi}}},
  \bibinfo{author}{\bibfnamefont{A.}~\bibnamefont{{Ferrara}}},
  \bibnamefont{and}
  \bibinfo{author}{\bibfnamefont{C.}~\bibnamefont{{Baccigalupi}}},
  \bibinfo{journal}{\mnras} \textbf{\bibinfo{volume}{360}},
  \bibinfo{pages}{1063} (\bibinfo{year}{2005}), \eprint{astro-ph/0502419}.

\bibitem[{\citenamefont{{Alvarez} et~al.}(2006)\citenamefont{{Alvarez},
  {Komatsu}, {Dor{\'e}}, and {Shapiro}}}]{2006ApJ...647..840A}
\bibinfo{author}{\bibfnamefont{M.~A.} \bibnamefont{{Alvarez}}},
  \bibinfo{author}{\bibfnamefont{E.}~\bibnamefont{{Komatsu}}},
  \bibinfo{author}{\bibfnamefont{O.}~\bibnamefont{{Dor{\'e}}}},
  \bibnamefont{and} \bibinfo{author}{\bibfnamefont{P.~R.}
  \bibnamefont{{Shapiro}}}, \bibinfo{journal}{\apj}
  \textbf{\bibinfo{volume}{647}}, \bibinfo{pages}{840} (\bibinfo{year}{2006}),
  \eprint{astro-ph/0512010}.

\bibitem[{\citenamefont{{Dor{\'e}} et~al.}(2004)\citenamefont{{Dor{\'e}},
  {Hennawi}, and {Spergel}}}]{2004ApJ...606...46D}
\bibinfo{author}{\bibfnamefont{O.}~\bibnamefont{{Dor{\'e}}}},
  \bibinfo{author}{\bibfnamefont{J.~F.} \bibnamefont{{Hennawi}}},
  \bibnamefont{and} \bibinfo{author}{\bibfnamefont{D.~N.}
  \bibnamefont{{Spergel}}}, \bibinfo{journal}{\apj}
  \textbf{\bibinfo{volume}{606}}, \bibinfo{pages}{46} (\bibinfo{year}{2004}),
  \eprint{astro-ph/0309337}.

\bibitem[{\citenamefont{{Ferraro} et~al.}(2015)\citenamefont{{Ferraro},
  {Sherwin}, and {Spergel}}}]{2015PhRvD..91h3533F}
\bibinfo{author}{\bibfnamefont{S.}~\bibnamefont{{Ferraro}}},
  \bibinfo{author}{\bibfnamefont{B.~D.} \bibnamefont{{Sherwin}}},
  \bibnamefont{and} \bibinfo{author}{\bibfnamefont{D.~N.}
  \bibnamefont{{Spergel}}}, \bibinfo{journal}{\prd}
  \textbf{\bibinfo{volume}{91}}, \bibinfo{eid}{083533} (\bibinfo{year}{2015}),
  \eprint{1401.1193}.

\bibitem[{\citenamefont{{Hu} and {Dodelson}}(2002)}]{2002ARA&A..40..171H}
\bibinfo{author}{\bibfnamefont{W.}~\bibnamefont{{Hu}}} \bibnamefont{and}
  \bibinfo{author}{\bibfnamefont{S.}~\bibnamefont{{Dodelson}}},
  \bibinfo{journal}{\araa} \textbf{\bibinfo{volume}{40}}, \bibinfo{pages}{171}
  (\bibinfo{year}{2002}), \eprint{astro-ph/0110414}.

\bibitem[{\citenamefont{{George} et~al.}(2015)\citenamefont{{George},
  {Reichardt}, {Aird}, {Benson}, {Bleem}, {Carlstrom}, {Chang}, {Cho},
  {Crawford}, {Crites} et~al.}}]{2015ApJ...799..177G}
\bibinfo{author}{\bibfnamefont{E.~M.} \bibnamefont{{George}}},
  \bibinfo{author}{\bibfnamefont{C.~L.} \bibnamefont{{Reichardt}}},
  \bibinfo{author}{\bibfnamefont{K.~A.} \bibnamefont{{Aird}}},
  \bibinfo{author}{\bibfnamefont{B.~A.} \bibnamefont{{Benson}}},
  \bibinfo{author}{\bibfnamefont{L.~E.} \bibnamefont{{Bleem}}},
  \bibinfo{author}{\bibfnamefont{J.~E.} \bibnamefont{{Carlstrom}}},
  \bibinfo{author}{\bibfnamefont{C.~L.} \bibnamefont{{Chang}}},
  \bibinfo{author}{\bibfnamefont{H.-M.} \bibnamefont{{Cho}}},
  \bibinfo{author}{\bibfnamefont{T.~M.} \bibnamefont{{Crawford}}},
  \bibinfo{author}{\bibfnamefont{A.~T.} \bibnamefont{{Crites}}},
  \bibnamefont{et~al.}, \bibinfo{journal}{\apj} \textbf{\bibinfo{volume}{799}},
  \bibinfo{eid}{177} (\bibinfo{year}{2015}), \eprint{1408.3161}.

\bibitem[{\citenamefont{{Planck Collaboration}
  et~al.}(2016)\citenamefont{{Planck Collaboration}, {Adam}, {Aghanim},
  {Ashdown}, {Aumont}, {Baccigalupi}, {Ballardini}, {Banday}, {Barreiro},
  {Bartolo} et~al.}}]{2016A&A...596A.108P}
\bibinfo{author}{\bibnamefont{{Planck Collaboration}}},
  \bibinfo{author}{\bibfnamefont{R.}~\bibnamefont{{Adam}}},
  \bibinfo{author}{\bibfnamefont{N.}~\bibnamefont{{Aghanim}}},
  \bibinfo{author}{\bibfnamefont{M.}~\bibnamefont{{Ashdown}}},
  \bibinfo{author}{\bibfnamefont{J.}~\bibnamefont{{Aumont}}},
  \bibinfo{author}{\bibfnamefont{C.}~\bibnamefont{{Baccigalupi}}},
  \bibinfo{author}{\bibfnamefont{M.}~\bibnamefont{{Ballardini}}},
  \bibinfo{author}{\bibfnamefont{A.~J.} \bibnamefont{{Banday}}},
  \bibinfo{author}{\bibfnamefont{R.~B.} \bibnamefont{{Barreiro}}},
  \bibinfo{author}{\bibfnamefont{N.}~\bibnamefont{{Bartolo}}},
  \bibnamefont{et~al.}, \bibinfo{journal}{\aap} \textbf{\bibinfo{volume}{596}},
  \bibinfo{eid}{A108} (\bibinfo{year}{2016}), \eprint{1605.03507}.

\bibitem[{\citenamefont{{Calabrese} et~al.}(2014)\citenamefont{{Calabrese},
  {Hlo{\v z}ek}, {Battaglia}, {Bond}, {de Bernardis}, {Devlin}, {Hajian},
  {Henderson}, {Hil}, {Kosowsky} et~al.}}]{2014JCAP...08..010C}
\bibinfo{author}{\bibfnamefont{E.}~\bibnamefont{{Calabrese}}},
  \bibinfo{author}{\bibfnamefont{R.}~\bibnamefont{{Hlo{\v z}ek}}},
  \bibinfo{author}{\bibfnamefont{N.}~\bibnamefont{{Battaglia}}},
  \bibinfo{author}{\bibfnamefont{J.~R.} \bibnamefont{{Bond}}},
  \bibinfo{author}{\bibfnamefont{F.}~\bibnamefont{{de Bernardis}}},
  \bibinfo{author}{\bibfnamefont{M.~J.} \bibnamefont{{Devlin}}},
  \bibinfo{author}{\bibfnamefont{A.}~\bibnamefont{{Hajian}}},
  \bibinfo{author}{\bibfnamefont{S.}~\bibnamefont{{Henderson}}},
  \bibinfo{author}{\bibfnamefont{J.~C.} \bibnamefont{{Hil}}},
  \bibinfo{author}{\bibfnamefont{A.}~\bibnamefont{{Kosowsky}}},
  \bibnamefont{et~al.}, \bibinfo{journal}{JCAP} \textbf{\bibinfo{volume}{8}},
  \bibinfo{eid}{010} (\bibinfo{year}{2014}), \eprint{1406.4794}.

\bibitem[{\citenamefont{{Smith} and {Ferraro}}(2016)}]{2016arXiv160701769S}
\bibinfo{author}{\bibfnamefont{K.~M.} \bibnamefont{{Smith}}} \bibnamefont{and}
  \bibinfo{author}{\bibfnamefont{S.}~\bibnamefont{{Ferraro}}},
  \bibinfo{journal}{ArXiv e-prints}  (\bibinfo{year}{2016}),
  \eprint{1607.01769}.

\bibitem[{\citenamefont{{Zahn} et~al.}(2012)\citenamefont{{Zahn}, {Reichardt},
  {Shaw}, {Lidz}, {Aird}, {Benson}, {Bleem}, {Carlstrom}, {Chang}, {Cho}
  et~al.}}]{2012ApJ...756...65Z}
\bibinfo{author}{\bibfnamefont{O.}~\bibnamefont{{Zahn}}},
  \bibinfo{author}{\bibfnamefont{C.~L.} \bibnamefont{{Reichardt}}},
  \bibinfo{author}{\bibfnamefont{L.}~\bibnamefont{{Shaw}}},
  \bibinfo{author}{\bibfnamefont{A.}~\bibnamefont{{Lidz}}},
  \bibinfo{author}{\bibfnamefont{K.~A.} \bibnamefont{{Aird}}},
  \bibinfo{author}{\bibfnamefont{B.~A.} \bibnamefont{{Benson}}},
  \bibinfo{author}{\bibfnamefont{L.~E.} \bibnamefont{{Bleem}}},
  \bibinfo{author}{\bibfnamefont{J.~E.} \bibnamefont{{Carlstrom}}},
  \bibinfo{author}{\bibfnamefont{C.~L.} \bibnamefont{{Chang}}},
  \bibinfo{author}{\bibfnamefont{H.~M.} \bibnamefont{{Cho}}},
  \bibnamefont{et~al.}, \bibinfo{journal}{\apj} \textbf{\bibinfo{volume}{756}},
  \bibinfo{eid}{65} (\bibinfo{year}{2012}), \eprint{1111.6386}.

\bibitem[{\citenamefont{{Mesinger} et~al.}(2012)\citenamefont{{Mesinger},
  {McQuinn}, and {Spergel}}}]{2012MNRAS.422.1403M}
\bibinfo{author}{\bibfnamefont{A.}~\bibnamefont{{Mesinger}}},
  \bibinfo{author}{\bibfnamefont{M.}~\bibnamefont{{McQuinn}}},
  \bibnamefont{and} \bibinfo{author}{\bibfnamefont{D.~N.}
  \bibnamefont{{Spergel}}}, \bibinfo{journal}{\mnras}
  \textbf{\bibinfo{volume}{422}}, \bibinfo{pages}{1403} (\bibinfo{year}{2012}),
  \eprint{1112.1820}.

\bibitem[{\citenamefont{{Battaglia} et~al.}(2013)\citenamefont{{Battaglia},
  {Natarajan}, {Trac}, {Cen}, and {Loeb}}}]{2013ApJ...776...83B}
\bibinfo{author}{\bibfnamefont{N.}~\bibnamefont{{Battaglia}}},
  \bibinfo{author}{\bibfnamefont{A.}~\bibnamefont{{Natarajan}}},
  \bibinfo{author}{\bibfnamefont{H.}~\bibnamefont{{Trac}}},
  \bibinfo{author}{\bibfnamefont{R.}~\bibnamefont{{Cen}}}, \bibnamefont{and}
  \bibinfo{author}{\bibfnamefont{A.}~\bibnamefont{{Loeb}}},
  \bibinfo{journal}{\apj} \textbf{\bibinfo{volume}{776}}, \bibinfo{eid}{83}
  (\bibinfo{year}{2013}), \eprint{1211.2832}.

\bibitem[{\citenamefont{{Park} et~al.}(2013)\citenamefont{{Park}, {Shapiro},
  {Komatsu}, {Iliev}, {Ahn}, and {Mellema}}}]{2013ApJ...769...93P}
\bibinfo{author}{\bibfnamefont{H.}~\bibnamefont{{Park}}},
  \bibinfo{author}{\bibfnamefont{P.~R.} \bibnamefont{{Shapiro}}},
  \bibinfo{author}{\bibfnamefont{E.}~\bibnamefont{{Komatsu}}},
  \bibinfo{author}{\bibfnamefont{I.~T.} \bibnamefont{{Iliev}}},
  \bibinfo{author}{\bibfnamefont{K.}~\bibnamefont{{Ahn}}}, \bibnamefont{and}
  \bibinfo{author}{\bibfnamefont{G.}~\bibnamefont{{Mellema}}},
  \bibinfo{journal}{\apj} \textbf{\bibinfo{volume}{769}}, \bibinfo{eid}{93}
  (\bibinfo{year}{2013}), \eprint{1301.3607}.

\bibitem[{\citenamefont{{Alvarez} and {Abel}}(2012)}]{2012ApJ...747..126A}
\bibinfo{author}{\bibfnamefont{M.~A.} \bibnamefont{{Alvarez}}}
  \bibnamefont{and} \bibinfo{author}{\bibfnamefont{T.}~\bibnamefont{{Abel}}},
  \bibinfo{journal}{\apj} \textbf{\bibinfo{volume}{747}}, \bibinfo{eid}{126}
  (\bibinfo{year}{2012}), \eprint{1003.6132}.

\bibitem[{\citenamefont{{Nelson} et~al.}(2015)\citenamefont{{Nelson},
  {Pillepich}, {Genel}, {Vogelsberger}, {Springel}, {Torrey},
  {Rodriguez-Gomez}, {Sijacki}, {Snyder}, {Griffen}
  et~al.}}]{2015A&C....13...12N}
\bibinfo{author}{\bibfnamefont{D.}~\bibnamefont{{Nelson}}},
  \bibinfo{author}{\bibfnamefont{A.}~\bibnamefont{{Pillepich}}},
  \bibinfo{author}{\bibfnamefont{S.}~\bibnamefont{{Genel}}},
  \bibinfo{author}{\bibfnamefont{M.}~\bibnamefont{{Vogelsberger}}},
  \bibinfo{author}{\bibfnamefont{V.}~\bibnamefont{{Springel}}},
  \bibinfo{author}{\bibfnamefont{P.}~\bibnamefont{{Torrey}}},
  \bibinfo{author}{\bibfnamefont{V.}~\bibnamefont{{Rodriguez-Gomez}}},
  \bibinfo{author}{\bibfnamefont{D.}~\bibnamefont{{Sijacki}}},
  \bibinfo{author}{\bibfnamefont{G.~F.} \bibnamefont{{Snyder}}},
  \bibinfo{author}{\bibfnamefont{B.}~\bibnamefont{{Griffen}}},
  \bibnamefont{et~al.}, \bibinfo{journal}{Astronomy and Computing}
  \textbf{\bibinfo{volume}{13}}, \bibinfo{pages}{12} (\bibinfo{year}{2015}),
  \eprint{1504.00362}.

\bibitem[{\citenamefont{{Vogelsberger}
  et~al.}(2013)\citenamefont{{Vogelsberger}, {Genel}, {Sijacki}, {Torrey},
  {Springel}, and {Hernquist}}}]{2013MNRAS.436.3031V}
\bibinfo{author}{\bibfnamefont{M.}~\bibnamefont{{Vogelsberger}}},
  \bibinfo{author}{\bibfnamefont{S.}~\bibnamefont{{Genel}}},
  \bibinfo{author}{\bibfnamefont{D.}~\bibnamefont{{Sijacki}}},
  \bibinfo{author}{\bibfnamefont{P.}~\bibnamefont{{Torrey}}},
  \bibinfo{author}{\bibfnamefont{V.}~\bibnamefont{{Springel}}},
  \bibnamefont{and}
  \bibinfo{author}{\bibfnamefont{L.}~\bibnamefont{{Hernquist}}},
  \bibinfo{journal}{\mnras} \textbf{\bibinfo{volume}{436}},
  \bibinfo{pages}{3031} (\bibinfo{year}{2013}), \eprint{1305.2913}.

\bibitem[{\citenamefont{{Ostriker} and {Vishniac}}(1986)}]{1986ApJ...306L..51O}
\bibinfo{author}{\bibfnamefont{J.~P.} \bibnamefont{{Ostriker}}}
  \bibnamefont{and} \bibinfo{author}{\bibfnamefont{E.~T.}
  \bibnamefont{{Vishniac}}}, \bibinfo{journal}{\apjl}
  \textbf{\bibinfo{volume}{306}}, \bibinfo{pages}{L51} (\bibinfo{year}{1986}).

\bibitem[{\citenamefont{{Vishniac}}(1987)}]{1987ApJ...322..597V}
\bibinfo{author}{\bibfnamefont{E.~T.} \bibnamefont{{Vishniac}}},
  \bibinfo{journal}{\apj} \textbf{\bibinfo{volume}{322}}, \bibinfo{pages}{597}
  (\bibinfo{year}{1987}).

\bibitem[{\citenamefont{{Dodelson} and {Jubas}}(1995)}]{1995ApJ...439..503D}
\bibinfo{author}{\bibfnamefont{S.}~\bibnamefont{{Dodelson}}} \bibnamefont{and}
  \bibinfo{author}{\bibfnamefont{J.~M.} \bibnamefont{{Jubas}}},
  \bibinfo{journal}{\apj} \textbf{\bibinfo{volume}{439}}, \bibinfo{pages}{503}
  (\bibinfo{year}{1995}), \eprint{astro-ph/9308019}.

\bibitem[{\citenamefont{{Jaffe} and
  {Kamionkowski}}(1998)}]{1998PhRvD..58d3001J}
\bibinfo{author}{\bibfnamefont{A.~H.} \bibnamefont{{Jaffe}}} \bibnamefont{and}
  \bibinfo{author}{\bibfnamefont{M.}~\bibnamefont{{Kamionkowski}}},
  \bibinfo{journal}{\prd} \textbf{\bibinfo{volume}{58}}, \bibinfo{eid}{043001}
  (\bibinfo{year}{1998}), \eprint{astro-ph/9801022}.

\bibitem[{\citenamefont{{Hu}}(2000)}]{2000ApJ...529...12H}
\bibinfo{author}{\bibfnamefont{W.}~\bibnamefont{{Hu}}}, \bibinfo{journal}{\apj}
  \textbf{\bibinfo{volume}{529}}, \bibinfo{pages}{12} (\bibinfo{year}{2000}),
  \eprint{astro-ph/9907103}.

\bibitem[{\citenamefont{{Ma} and {Fry}}(2002)}]{2002PhRvL..88u1301M}
\bibinfo{author}{\bibfnamefont{C.-P.} \bibnamefont{{Ma}}} \bibnamefont{and}
  \bibinfo{author}{\bibfnamefont{J.~N.} \bibnamefont{{Fry}}},
  \bibinfo{journal}{Physical Review Letters} \textbf{\bibinfo{volume}{88}},
  \bibinfo{eid}{211301} (\bibinfo{year}{2002}), \eprint{astro-ph/0106342}.

\bibitem[{\citenamefont{{Zhang} et~al.}(2004)\citenamefont{{Zhang}, {Pen}, and
  {Trac}}}]{2004MNRAS.347.1224Z}
\bibinfo{author}{\bibfnamefont{P.}~\bibnamefont{{Zhang}}},
  \bibinfo{author}{\bibfnamefont{U.-L.} \bibnamefont{{Pen}}}, \bibnamefont{and}
  \bibinfo{author}{\bibfnamefont{H.}~\bibnamefont{{Trac}}},
  \bibinfo{journal}{\mnras} \textbf{\bibinfo{volume}{347}},
  \bibinfo{pages}{1224} (\bibinfo{year}{2004}), \eprint{astro-ph/0304534}.

\bibitem[{\citenamefont{{Shaw} et~al.}(2012)\citenamefont{{Shaw}, {Rudd}, and
  {Nagai}}}]{2012ApJ...756...15S}
\bibinfo{author}{\bibfnamefont{L.~D.} \bibnamefont{{Shaw}}},
  \bibinfo{author}{\bibfnamefont{D.~H.} \bibnamefont{{Rudd}}},
  \bibnamefont{and} \bibinfo{author}{\bibfnamefont{D.}~\bibnamefont{{Nagai}}},
  \bibinfo{journal}{\apj} \textbf{\bibinfo{volume}{756}}, \bibinfo{eid}{15}
  (\bibinfo{year}{2012}), \eprint{1109.0553}.

\bibitem[{\citenamefont{{Park} et~al.}(2016)\citenamefont{{Park}, {Komatsu},
  {Shapiro}, {Koda}, and {Mao}}}]{2016ApJ...818...37P}
\bibinfo{author}{\bibfnamefont{H.}~\bibnamefont{{Park}}},
  \bibinfo{author}{\bibfnamefont{E.}~\bibnamefont{{Komatsu}}},
  \bibinfo{author}{\bibfnamefont{P.~R.} \bibnamefont{{Shapiro}}},
  \bibinfo{author}{\bibfnamefont{J.}~\bibnamefont{{Koda}}}, \bibnamefont{and}
  \bibinfo{author}{\bibfnamefont{Y.}~\bibnamefont{{Mao}}},
  \bibinfo{journal}{\apj} \textbf{\bibinfo{volume}{818}}, \bibinfo{eid}{37}
  (\bibinfo{year}{2016}), \eprint{1506.05177}.

\bibitem[{\citenamefont{{Alvarez}}(2016)}]{2016ApJ...824..118A}
\bibinfo{author}{\bibfnamefont{M.~A.} \bibnamefont{{Alvarez}}},
  \bibinfo{journal}{\apj} \textbf{\bibinfo{volume}{824}}, \bibinfo{eid}{118}
  (\bibinfo{year}{2016}), \eprint{1511.02846}.

\bibitem[{\citenamefont{{Springel}}(2010)}]{2010MNRAS.401..791S}
\bibinfo{author}{\bibfnamefont{V.}~\bibnamefont{{Springel}}},
  \bibinfo{journal}{\mnras} \textbf{\bibinfo{volume}{401}},
  \bibinfo{pages}{791} (\bibinfo{year}{2010}), \eprint{0901.4107}.

\bibitem[{\citenamefont{{Torrey} et~al.}(2014)\citenamefont{{Torrey},
  {Vogelsberger}, {Genel}, {Sijacki}, {Springel}, and
  {Hernquist}}}]{2014MNRAS.438.1985T}
\bibinfo{author}{\bibfnamefont{P.}~\bibnamefont{{Torrey}}},
  \bibinfo{author}{\bibfnamefont{M.}~\bibnamefont{{Vogelsberger}}},
  \bibinfo{author}{\bibfnamefont{S.}~\bibnamefont{{Genel}}},
  \bibinfo{author}{\bibfnamefont{D.}~\bibnamefont{{Sijacki}}},
  \bibinfo{author}{\bibfnamefont{V.}~\bibnamefont{{Springel}}},
  \bibnamefont{and}
  \bibinfo{author}{\bibfnamefont{L.}~\bibnamefont{{Hernquist}}},
  \bibinfo{journal}{\mnras} \textbf{\bibinfo{volume}{438}},
  \bibinfo{pages}{1985} (\bibinfo{year}{2014}), \eprint{1305.4931}.

\bibitem[{\citenamefont{{Springel} et~al.}(2001)\citenamefont{{Springel},
  {White}, {Tormen}, and {Kauffmann}}}]{2001MNRAS.328..726S}
\bibinfo{author}{\bibfnamefont{V.}~\bibnamefont{{Springel}}},
  \bibinfo{author}{\bibfnamefont{S.~D.~M.} \bibnamefont{{White}}},
  \bibinfo{author}{\bibfnamefont{G.}~\bibnamefont{{Tormen}}}, \bibnamefont{and}
  \bibinfo{author}{\bibfnamefont{G.}~\bibnamefont{{Kauffmann}}},
  \bibinfo{journal}{\mnras} \textbf{\bibinfo{volume}{328}},
  \bibinfo{pages}{726} (\bibinfo{year}{2001}), \eprint{astro-ph/0012055}.

\bibitem[{\citenamefont{{Vogelsberger}
  et~al.}(2014)\citenamefont{{Vogelsberger}, {Genel}, {Springel}, {Torrey},
  {Sijacki}, {Xu}, {Snyder}, {Bird}, {Nelson}, and
  {Hernquist}}}]{2014Natur.509..177V}
\bibinfo{author}{\bibfnamefont{M.}~\bibnamefont{{Vogelsberger}}},
  \bibinfo{author}{\bibfnamefont{S.}~\bibnamefont{{Genel}}},
  \bibinfo{author}{\bibfnamefont{V.}~\bibnamefont{{Springel}}},
  \bibinfo{author}{\bibfnamefont{P.}~\bibnamefont{{Torrey}}},
  \bibinfo{author}{\bibfnamefont{D.}~\bibnamefont{{Sijacki}}},
  \bibinfo{author}{\bibfnamefont{D.}~\bibnamefont{{Xu}}},
  \bibinfo{author}{\bibfnamefont{G.}~\bibnamefont{{Snyder}}},
  \bibinfo{author}{\bibfnamefont{S.}~\bibnamefont{{Bird}}},
  \bibinfo{author}{\bibfnamefont{D.}~\bibnamefont{{Nelson}}}, \bibnamefont{and}
  \bibinfo{author}{\bibfnamefont{L.}~\bibnamefont{{Hernquist}}},
  \bibinfo{journal}{\nat} \textbf{\bibinfo{volume}{509}}, \bibinfo{pages}{177}
  (\bibinfo{year}{2014}), \eprint{1405.1418}.

\bibitem[{\citenamefont{{Genel} et~al.}(2014)\citenamefont{{Genel},
  {Vogelsberger}, {Springel}, {Sijacki}, {Nelson}, {Snyder}, {Rodriguez-Gomez},
  {Torrey}, and {Hernquist}}}]{2014MNRAS.445..175G}
\bibinfo{author}{\bibfnamefont{S.}~\bibnamefont{{Genel}}},
  \bibinfo{author}{\bibfnamefont{M.}~\bibnamefont{{Vogelsberger}}},
  \bibinfo{author}{\bibfnamefont{V.}~\bibnamefont{{Springel}}},
  \bibinfo{author}{\bibfnamefont{D.}~\bibnamefont{{Sijacki}}},
  \bibinfo{author}{\bibfnamefont{D.}~\bibnamefont{{Nelson}}},
  \bibinfo{author}{\bibfnamefont{G.}~\bibnamefont{{Snyder}}},
  \bibinfo{author}{\bibfnamefont{V.}~\bibnamefont{{Rodriguez-Gomez}}},
  \bibinfo{author}{\bibfnamefont{P.}~\bibnamefont{{Torrey}}}, \bibnamefont{and}
  \bibinfo{author}{\bibfnamefont{L.}~\bibnamefont{{Hernquist}}},
  \bibinfo{journal}{\mnras} \textbf{\bibinfo{volume}{445}},
  \bibinfo{pages}{175} (\bibinfo{year}{2014}), \eprint{1405.3749}.

\bibitem[{\citenamefont{{Smith} et~al.}(2003)\citenamefont{{Smith}, {Peacock},
  {Jenkins}, {White}, {Frenk}, {Pearce}, {Thomas}, {Efstathiou}, and
  {Couchman}}}]{2003MNRAS.341.1311S}
\bibinfo{author}{\bibfnamefont{R.~E.} \bibnamefont{{Smith}}},
  \bibinfo{author}{\bibfnamefont{J.~A.} \bibnamefont{{Peacock}}},
  \bibinfo{author}{\bibfnamefont{A.}~\bibnamefont{{Jenkins}}},
  \bibinfo{author}{\bibfnamefont{S.~D.~M.} \bibnamefont{{White}}},
  \bibinfo{author}{\bibfnamefont{C.~S.} \bibnamefont{{Frenk}}},
  \bibinfo{author}{\bibfnamefont{F.~R.} \bibnamefont{{Pearce}}},
  \bibinfo{author}{\bibfnamefont{P.~A.} \bibnamefont{{Thomas}}},
  \bibinfo{author}{\bibfnamefont{G.}~\bibnamefont{{Efstathiou}}},
  \bibnamefont{and} \bibinfo{author}{\bibfnamefont{H.~M.~P.}
  \bibnamefont{{Couchman}}}, \bibinfo{journal}{\mnras}
  \textbf{\bibinfo{volume}{341}}, \bibinfo{pages}{1311} (\bibinfo{year}{2003}),
  \eprint{astro-ph/0207664}.

\bibitem[{\citenamefont{{Bond}}(1996)}]{1996LesHouches}
\bibinfo{author}{\bibfnamefont{J.~R.} \bibnamefont{{Bond}}},
  \bibinfo{journal}{``Cosmology and Large Scale Structure'', Les Houches
  Session LX, August 1993, ed. R. Schaeffer, Elsevier Science Press} pp.
  \bibinfo{pages}{469--674} (\bibinfo{year}{1996}).

\bibitem[{\citenamefont{{Bond} and {Myers}}(1996)}]{1996ApJS..103...63B}
\bibinfo{author}{\bibfnamefont{J.~R.} \bibnamefont{{Bond}}} \bibnamefont{and}
  \bibinfo{author}{\bibfnamefont{S.~T.} \bibnamefont{{Myers}}},
  \bibinfo{journal}{\apjs} \textbf{\bibinfo{volume}{103}}, \bibinfo{pages}{63}
  (\bibinfo{year}{1996}).

\bibitem[{\citenamefont{{Alvarez} et~al.}(2017)\citenamefont{{Alvarez}, {Bond},
  {Stein}, {Bahmanyar}, {Battaglia}, {Hajian}, {Pham}, and {van
  Engelen}}}]{2017abs}
\bibinfo{author}{\bibfnamefont{M.~A.} \bibnamefont{{Alvarez}}},
  \bibinfo{author}{\bibfnamefont{J.~R.} \bibnamefont{{Bond}}},
  \bibinfo{author}{\bibfnamefont{G.}~\bibnamefont{{Stein}}},
  \bibinfo{author}{\bibfnamefont{A.}~\bibnamefont{{Bahmanyar}}},
  \bibinfo{author}{\bibfnamefont{N.}~\bibnamefont{{Battaglia}}},
  \bibinfo{author}{\bibfnamefont{A.}~\bibnamefont{{Hajian}}},
  \bibinfo{author}{\bibfnamefont{L.}~\bibnamefont{{Pham}}}, \bibnamefont{and}
  \bibinfo{author}{\bibfnamefont{A.}~\bibnamefont{{van Engelen}}},
  \bibinfo{journal}{in preparation}  (\bibinfo{year}{2017}).

\bibitem[{\citenamefont{{Battaglia} et~al.}(2012)\citenamefont{{Battaglia},
  {Bond}, {Pfrommer}, and {Sievers}}}]{2012ApJ...758...75B}
\bibinfo{author}{\bibfnamefont{N.}~\bibnamefont{{Battaglia}}},
  \bibinfo{author}{\bibfnamefont{J.~R.} \bibnamefont{{Bond}}},
  \bibinfo{author}{\bibfnamefont{C.}~\bibnamefont{{Pfrommer}}},
  \bibnamefont{and} \bibinfo{author}{\bibfnamefont{J.~L.}
  \bibnamefont{{Sievers}}}, \bibinfo{journal}{\apj}
  \textbf{\bibinfo{volume}{758}}, \bibinfo{eid}{75} (\bibinfo{year}{2012}),
  \eprint{1109.3711}.

\bibitem[{\citenamefont{{Efstathiou} and {Bond}}(1987)}]{1987MNRAS.227P..33E}
\bibinfo{author}{\bibfnamefont{G.}~\bibnamefont{{Efstathiou}}}
  \bibnamefont{and} \bibinfo{author}{\bibfnamefont{J.~R.}
  \bibnamefont{{Bond}}}, \bibinfo{journal}{\mnras}
  \textbf{\bibinfo{volume}{227}}, \bibinfo{pages}{33P} (\bibinfo{year}{1987}).

\bibitem[{\citenamefont{{Trac} et~al.}(2011)\citenamefont{{Trac}, {Bode}, and
  {Ostriker}}}]{2011ApJ...727...94T}
\bibinfo{author}{\bibfnamefont{H.}~\bibnamefont{{Trac}}},
  \bibinfo{author}{\bibfnamefont{P.}~\bibnamefont{{Bode}}}, \bibnamefont{and}
  \bibinfo{author}{\bibfnamefont{J.~P.} \bibnamefont{{Ostriker}}},
  \bibinfo{journal}{\apj} \textbf{\bibinfo{volume}{727}}, \bibinfo{eid}{94}
  (\bibinfo{year}{2011}), \eprint{1006.2828}.

\bibitem[{\citenamefont{{Battaglia} et~al.}(2010)\citenamefont{{Battaglia},
  {Bond}, {Pfrommer}, {Sievers}, and {Sijacki}}}]{2010ApJ...725...91B}
\bibinfo{author}{\bibfnamefont{N.}~\bibnamefont{{Battaglia}}},
  \bibinfo{author}{\bibfnamefont{J.~R.} \bibnamefont{{Bond}}},
  \bibinfo{author}{\bibfnamefont{C.}~\bibnamefont{{Pfrommer}}},
  \bibinfo{author}{\bibfnamefont{J.~L.} \bibnamefont{{Sievers}}},
  \bibnamefont{and}
  \bibinfo{author}{\bibfnamefont{D.}~\bibnamefont{{Sijacki}}},
  \bibinfo{journal}{\apj} \textbf{\bibinfo{volume}{725}}, \bibinfo{pages}{91}
  (\bibinfo{year}{2010}), \eprint{1003.4256}.

\bibitem[{\citenamefont{{Dolag} et~al.}(2016)\citenamefont{{Dolag}, {Komatsu},
  and {Sunyaev}}}]{2016MNRAS.463.1797D}
\bibinfo{author}{\bibfnamefont{K.}~\bibnamefont{{Dolag}}},
  \bibinfo{author}{\bibfnamefont{E.}~\bibnamefont{{Komatsu}}},
  \bibnamefont{and}
  \bibinfo{author}{\bibfnamefont{R.}~\bibnamefont{{Sunyaev}}},
  \bibinfo{journal}{\mnras} \textbf{\bibinfo{volume}{463}},
  \bibinfo{pages}{1797} (\bibinfo{year}{2016}), \eprint{1509.05134}.

\bibitem[{\citenamefont{{Roncarelli} et~al.}(2017)\citenamefont{{Roncarelli},
  {Villaescusa-Navarro}, and {Baldi}}}]{2017MNRAS.tmp..176R}
\bibinfo{author}{\bibfnamefont{M.}~\bibnamefont{{Roncarelli}}},
  \bibinfo{author}{\bibfnamefont{F.}~\bibnamefont{{Villaescusa-Navarro}}},
  \bibnamefont{and} \bibinfo{author}{\bibfnamefont{M.}~\bibnamefont{{Baldi}}},
  \bibinfo{journal}{\mnras}  (\bibinfo{year}{2017}), \eprint{1702.00676}.

\bibitem[{\citenamefont{{Iliev} et~al.}(2007)\citenamefont{{Iliev}, {Pen},
  {Bond}, {Mellema}, and {Shapiro}}}]{2007ApJ...660..933I}
\bibinfo{author}{\bibfnamefont{I.~T.} \bibnamefont{{Iliev}}},
  \bibinfo{author}{\bibfnamefont{U.-L.} \bibnamefont{{Pen}}},
  \bibinfo{author}{\bibfnamefont{J.~R.} \bibnamefont{{Bond}}},
  \bibinfo{author}{\bibfnamefont{G.}~\bibnamefont{{Mellema}}},
  \bibnamefont{and} \bibinfo{author}{\bibfnamefont{P.~R.}
  \bibnamefont{{Shapiro}}}, \bibinfo{journal}{\apj}
  \textbf{\bibinfo{volume}{660}}, \bibinfo{pages}{933} (\bibinfo{year}{2007}),
  \eprint{astro-ph/0609592}.

\end{thebibliography}

\appendix

\section{CORRECTION FOR THE MISSING POWER IN THE SIMULATION BOX} \label{app:missing power}

\begin{figure}
  \begin{center}
    \includegraphics[width=0.45\textwidth]{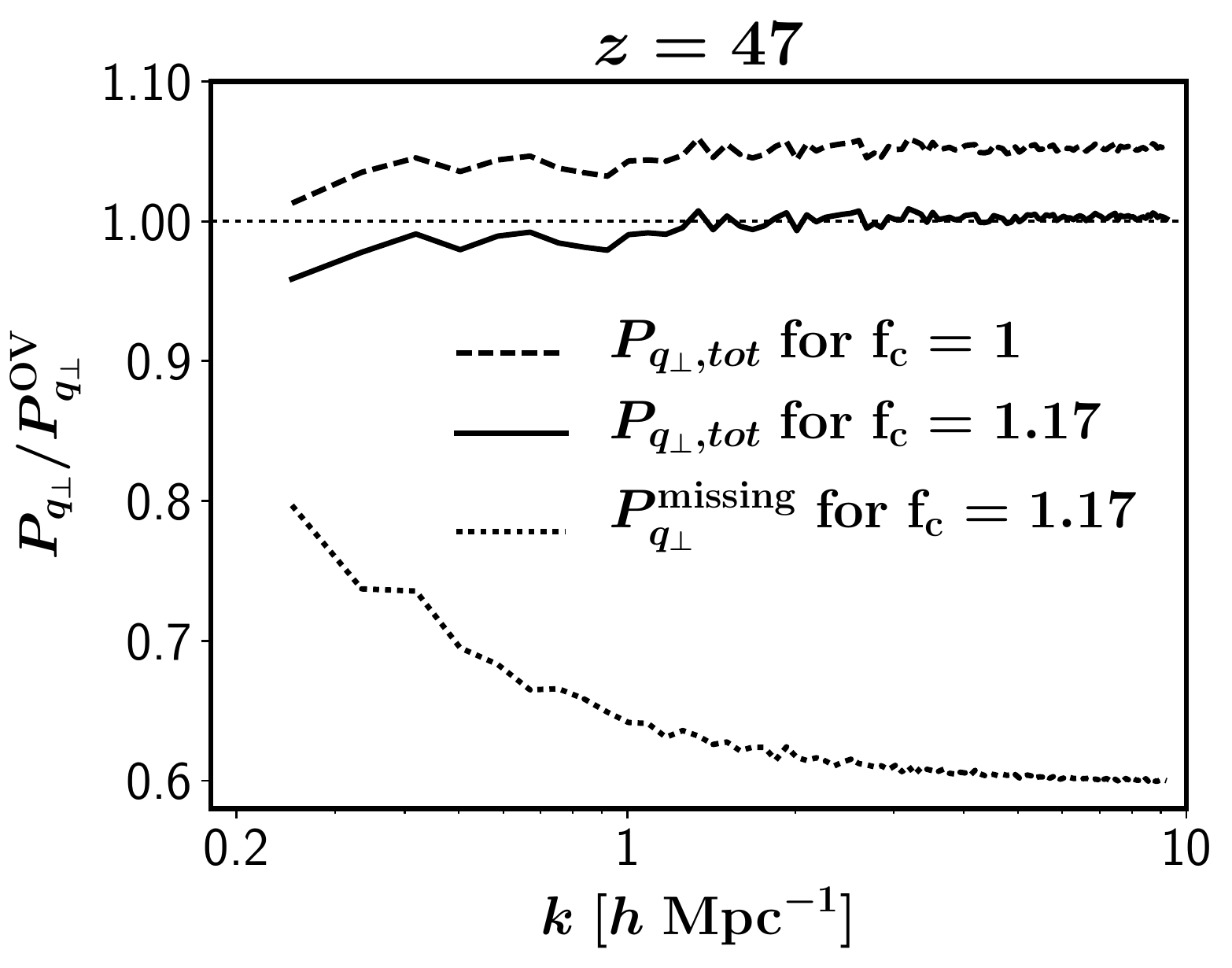}
    \includegraphics[width=0.45\textwidth]{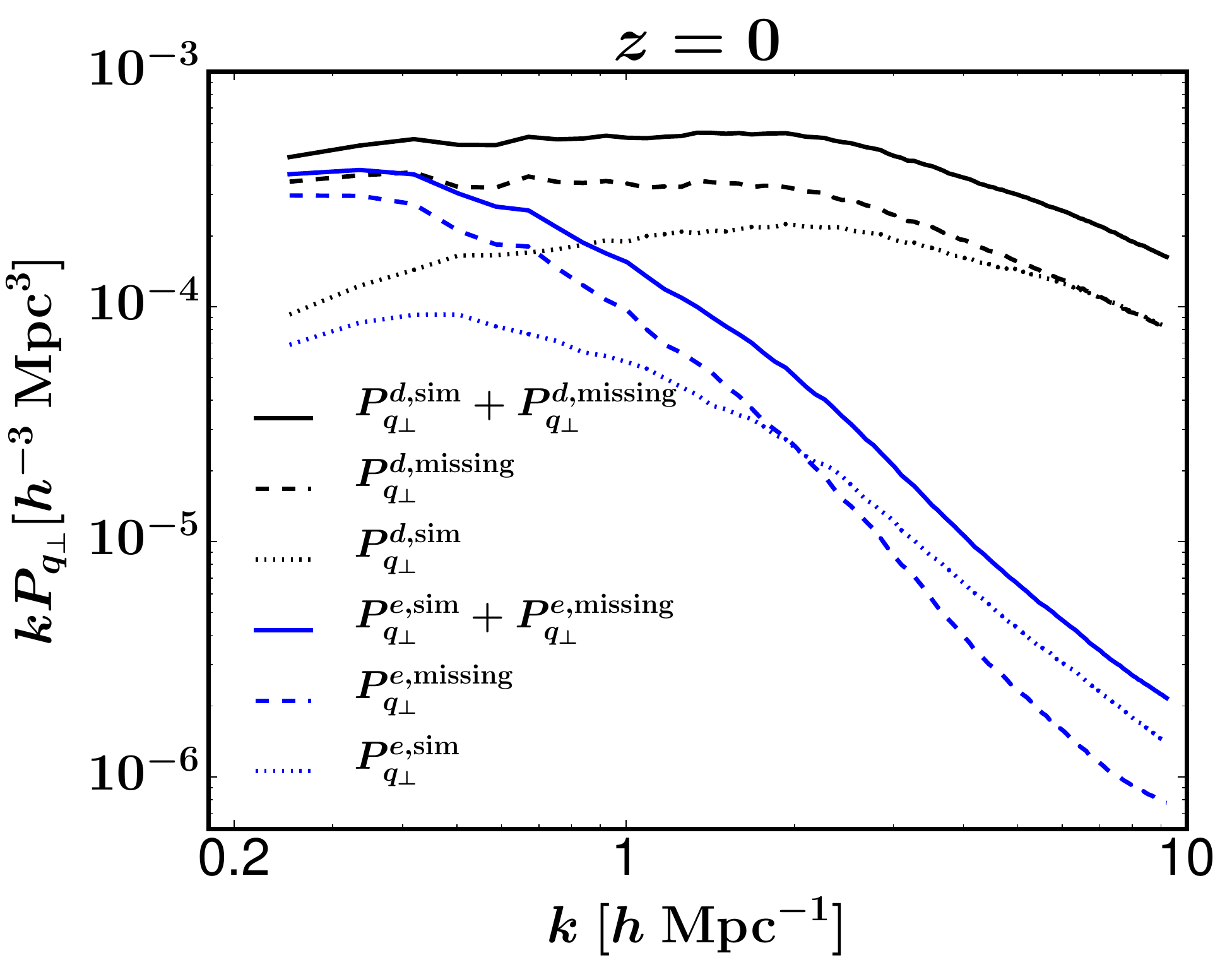}
 \caption{{\em left} -- Ratio between the missing-power-corrected transverse momentum power spectrum at $z=47$ and the corresponding Ostriker-Vishniac spectrum. The dashed line describes the case that uses $k_{\rm box}$ in the truncation of the wavenumber in the missing power correction while the solid line uses $(1.17)k_{\rm box}$. {\em right} -- $P^{\rm sim}_{q_\perp}$ (dotted), $P^{\rm missing}_{q_\perp}$ (dashed), and their sum (solid) multiplied by the wavenumber $k$ as functions of $k$ for dark matter (black) and free electron (blue) at $z=0$.}
 \label{fig:appendix_missing_power}
  \end{center}
\end{figure}

As shown in Equation~(\ref{eq:Pq_unconnected}), $P_{q_\perp}$ is given by a convolution of $P_{\delta\delta}$ and $P_{vv}$. This means $P_{q_\perp}$ at a single wavenumber would receive contributions from $P_{\delta\delta}$ and $P_{vv}$ at a range of wavenumbers. Failing to sample density and velocity mode for the whole range of wavenumbers will thus result in underestimation of $P_{q_\perp}$. In simulations, finite box-size truncate the long wave contributions and can cause this problem \citep{2004MNRAS.347.1224Z, 2007ApJ...660..933I,2013ApJ...769...93P}:   the integral  in Equation~(\ref{eq:Pq_unconnected}) is truncated for wavelengths longer than the size of the box, $l_{\rm box} = 75~h^{-1}~{\rm Mpc}$.

In previous works, the truncation was often set to the fundamental mode $k_{\rm box}\equiv 2\pi/l_{\rm box}$ of the box and the missing power, $P^{\rm missing}_{q_\perp}$, was then estimated by evaluating the integral for $k^\prime<k_{\rm box}$ or $|\mathbf{k}-\mathbf{k}^\prime|<k_{\rm box}$,  assuming linear $P_{\delta\delta}$ and $P_{vv}$ at $k<k_{\rm box}$. We have tested this method for the $z=47$ snapshot, where the density and velocity fluctuations should be in the linear regime. If the correction is accurate we are expected to recover Equation~(\ref{eq:Pq_unconnected}) evaluated with linear density and velocity power spectra, which is known as the Ostriker-Vishniac spectrum \citep[$P^{\rm OV}_{q_\perp}$;][]{1987MNRAS.227P..33E,1987ApJ...322..597V}. The result shown as the dashed line in Figure~\ref{fig:appendix_missing_power} indicates that this correction method overshoots the correct (measured in simulation) value by $\sim5\%$.

 We can improve the correction further by adjusting the spherical truncation wavenumber with a fudge factor, $f_c$,  multiplying the reciprocal lattice cubic size $k_{\rm box}$. An estimate of $f_c$ is obtained by equating the volume in a spherical octant out to $f_c k_{\rm box}$ to the reciprocal lattice volume, $f_c =(6/\pi)^{1/3} \sim 1.2$, with refinements depending upon the shape of the power being integrated. We find that when we truncated the integral at $f_c k_{\rm box}$ with $f_c=1.17$, we can suppress the error down to near a percent or less for most of the wavenumbers as described by the solid line in Figure~\ref{fig:appendix_missing_power}. We note that the best-fit value for $f_c$ may be different in other simulations with different box sizes.

We therefore use
\bea \label{eq:missing_power}
 P^{\rm missing}_{q_\perp}
 = 
 \int_{k^\prime<f_c k_{\rm box}~{\rm or }~|\mathbf{k}-\mathbf{k}^\prime|<f_c k_{\rm box}} \frac{d^3k^\prime}{(2\pi)^3} P_{\delta\delta} (|\mathbf{k} - \mathbf{k^\prime}|,z) P_{vv}(k^\prime,z)
\frac{k(k - 2k^\prime\mu^\prime)(1-{\mu^\prime}^2)}{k^2 + {k^\prime}^2-2kk^\prime\mu^\prime}
\eea
for the missing power. $P^{\rm sim}_{q_\perp}$, $P^{\rm missing}_{q_\perp}$, and their sum are shown in Figure~\ref{fig:appendix_missing_power}. For all the wavenumbers available, $P^{\rm missing}_{q_\perp}$ is at least comparable to and often larger than $P^{\rm sim}_{q_\perp}$ both for dark matter and electron density.

\end{document}